\documentstyle[prd,eqsecnum,aps,epsf]{revtex}

\def\be{\begin{equation}}
\def\ee{\end{equation}}
\def\ba{\begin{eqnarray}}
\def\ea{\end{eqnarray}}
\newcommand{\beqa}{\begin{eqnarray}}
\newcommand{\eeqa}{\end{eqnarray}}
\newcommand{\bea}{\begin{eqnarray}}
\newcommand{\eea}{\end{eqnarray}}

\newcommand{\singlefig}[2]{
\begin{center}
\begin{minipage}{#1}
\epsfxsize=#1
\epsffile{#2}
\end{minipage}
\end{center}}
\newenvironment{figcaption}[2]{
 \vspace{0.3cm}
 \refstepcounter{figure}
 \label{#1}
 \begin{center}
 \begin{minipage}{#2}
 \begingroup \small FIG. \thefigure: }{
 \endgroup
 \end{minipage}
 \end{center}}
\def\beq{\begin{equation}}
\def\eeq{\end{equation}}
\newcommand{\gsim}{\mbox{\raisebox{-1.ex}{$\stackrel
     {\textstyle>}{\textstyle\sim}$}}}
\newcommand{\lsim}{\mbox{\raisebox{-1.ex}{$\stackrel
     {\textstyle<}{\textstyle \sim}$}}}
\newcommand{\square}{\kern1pt\vbox{\hrule height
1.2pt\hbox{\vrule width 1.2pt\hskip 3pt
   \vbox{\vskip 6pt}\hskip 3pt\vrule width 0.6pt}\hrule
height 0.6pt}\kern1pt}

\begin{document}

\title{On the Construction of Nonsingular Pre-Big-Bang and Ekpyrotic \\
Cosmologies and the Resulting Density Perturbations} 

\author{Shinji Tsujikawa$^{1)}$, Robert Brandenberger$^{2)}$ 
and Fabio Finelli$^{3)}$ } 

\address{1) Research Center for the Early 
Universe, University of Tokyo, Hongo, Bunkyo-ku, Tokyo 169-8555, 
Japan.\\
email: shinji@resceu.s.u-tokyo.ac.jp\\
2) Institut d'Astrophysique de Paris, 98bis Blvd. Arago, F-75014 Paris,
France,
\\and\\
Physics Department, Brown University, Providence, RI 02912, USA.\\
email: rhb@het.brown.edu\\
3) I.A.S.F. - Sezione di Bologna, C.N.R., Via Gobetti 101, 40129 Bologna,
Italy.\\
email: finelli@tesre.bo.cnr.it\\
[.3em]} 
\date{\today} 
\maketitle
\begin{abstract}
We consider the construction of nonsingular Pre-Big-Bang and Ekpyrotic
type cosmological
models realized by the addition to the action of specific higher-order terms
stemming from quantum corrections. We study models involving
general relativity coupled to a single scalar field with a
potential motivated by the Ekpyrotic scenario. We find that the inclusion of 
the string loop and quantum correction terms in the string frame makes it 
possible to obtain solutions of the variational equations which are
nonsingular and bouncing in the Einstein frame, 
even when a negative exponential 
potential is present, as is the case in the Ekpyrotic scenario. This
allows us to discuss the evolution of cosmological perturbations
without the need to invoke matching conditions between
two Einstein Universes, one representing the contracting branch, the second
the expanding branch. We analyze the spectra of perturbations 
produced during the bouncing phase and find that the spectrum of 
curvature fluctuations
in the model proposed originally to implement the Ekpyrotic scenario
has a large blue tilt ($n_{\cal R}= 3$).  Except for instabilities introduced 
on small scales, the result agrees with what is obtained by imposing 
continuity of the induced metric and of the extrinsic curvature across a 
constant scalar field (up to $k^2$ corrections equal to the constant energy 
density) matching surface between the contracting and the expanding 
Einstein Universes.  We also discuss nonsingular cosmological solutions 
obtained when a Gauss-Bonnet term with a coefficient suitably dependent on 
the scalar matter field is added to the action in the Einstein frame with a 
potential for the scalar field present.  In this scenario, nonsingular 
solutions are found which start in an asymptotically flat state, 
undergo a period of super-exponential inflation and end with a graceful exit.  
The spectrum of fluctuations is also calculated in this case.
\end{abstract}

\vskip 1pc 
\vskip 2pc

\baselineskip = 14pt

\section{Introduction}                            

There has recently been a lot of interest in cosmological scenarios in
which it is assumed that instead of emerging from an initial Big Bang
singularity, our Universe has resulted from an Einstein frame 
bounce which connects a
previous contracting phase with the present phase of cosmological
expansion. A lot of this interest has been fueled by string cosmology,
the attempt to merge string theory and cosmology. Pre-Big-Bang (PBB)
cosmology \cite{Veneziano:1991ek,Gasperini:1992em} 
(see \cite{Lidsey:1999mc,Gasperini:2002} for a comprehensive review) and the 
Ekpyrotic scenario \cite{Khoury:2001wf} are two well-known
models in which our present phase of cosmological expansion is
postulated to have emerged from a previous phase of cosmological
contraction \footnote{In PBB cosmology this statement is true from the point 
of view of the Einstein frame metric, in the Ekpyrotic scenario it is true 
from the point of view of the four space-time dimensional effective action 
which is used to describe the cosmology.}.  In both examples, however, the 
cosmological description in terms of an effective action breaks down at the 
bounce.  In the case of PBB cosmology this bounce corresponds to a region of 
high curvature where higher derivative and string corrections to the 
effective action will be important, in the case of the Ekpyrotic scenario 
the bounce occurs when two four space-time dimensional branes collide in a 
five dimensional bulk.

Models with a cosmological bounce potentially provide an alternative
to cosmological inflation in addressing the homogeneity problem
of standard cosmology and in yielding a causal mechanism of structure
formation, the latter since at times long before the bounce fixed
comoving scales of cosmological interest today will have been inside
the Hubble radius.\footnote{See 
refs.~\cite{Kallosh:2001ai,Kallosh:2001du,Enqvist:2001zk,Rasanen:2001hf,Felder:2002jk,Linde:2002ws}
 for critical arguments on the Ekpyrotic scenario.}
However, since in both PBB and
Ekpyrotic scenarios the Hubble parameter increases during the
collapsing phase, symmetry arguments such as those used originally
\cite{Press} to predict the scale-invariance of cosmological fluctuations
in inflationary cosmology would lead one to expect a blue
spectrum of curvature perturbations in these models, at least in
effective field theory models in 
which there is only one ``matter'' field. As outlined in Appendix A, in
PBB cosmology one expects a spectrum with spectral index $n=4$,
whereas in the Ekpyrotic scenario one expects $n=3$. In the case of
PBB cosmology, this heuristic prediction was confirmed 
\cite{BGGMV} by a general relativistic analysis (which is, however, 
subject to the caveats indicated below). In the case of the Ekpyrotic
scenario, there is a large disagreement in the results. Whereas
the works of 
\cite{Lyth:2001pf,Brandenberger:2001bs,Hwang:2001ga,Tsujikawa:2001ad} 
yield results in
agreement with the heuristic prediction (namely $n=3$), others
\cite{Khoury:2001zk,Durrer:2002jn} obtain a scale-invariant 
spectrum of adiabatic fluctuations (thus also casting
doubt on past results in the literature on the spectrum of
fluctuations in the PBB scenario).

The singularity of the effective action at the time of the bounce
makes it impossible to follow the evolution of the background
cosmology and of the resulting cosmological perturbations
rigorously \cite{Lyth:2001nv}. 
In much of the previous work, both in the context of PBB 
cosmology \cite{Deruelle:1995kd} and of the Ekpyrotic scenario 
\cite{Brandenberger:2001bs,Hwang:2001ga,Tsujikawa:2001ad,Durrer:2002jn}, 
the fluctuations were computed by
matching two Einstein Universes (the first representing the contracting
phase, the second the expanding phase) along a space-like surface
(representing the bounce region) and applying continuity of the
induced metric and of the extrinsic curvature across the surface
\cite{Hwang:1991an,Deruelle:1995kd}. As emphasized in \cite{Durrer:2002jn}, 
the result will depend on
how the matching surface is chosen \footnote{As emphasized already
in \cite{Brandenberger:2001bs} and \cite{Finelli:2001sr}, there is a 
consistency
check for proposed matching surfaces: when applied to the reheating
surface in inflationary cosmology, the correct result should emerge.
This does not happen with the prescription advocated in 
\cite{Durrer:2002jn}, nor
does it with the matching prescription of \cite{Khoury:2001zk} which is
not based on a geometric analysis (see also 
\cite{Martin:2001ue,Martin:2002ar} for
a criticism of the latter matching prescription).}. 

In the context of PBB cosmology, it was realized 
\cite{Gasperini:1996fu,Brustein:1997cv,Brustein:1997xn} (see
also \cite{Rey:1996ad,Foffa:1999dv,Cartier:1999vk}) that
higher derivative corrections (defined in the string frame)
to the action induced by inverse
string tension and coupling constant corrections can yield a
nonsingular background cosmology
\footnote{Construction of nonsingular cosmologies in pure Einstein
gravity by means of specific higher derivative terms is also
possible (see e.g. \cite{MB,BMS} and its application to
PBB cosmology in \cite{Brandenberger:1998zs,Easson:1999xw}).}. 
This then allows the
study of the evolution of cosmological perturbations without having
to use ad hoc matching prescriptions. The effects of the higher
derivative terms in the action on the evolution of fluctuations in
the PBB cosmology was investigated in \cite{Cartier:2001is}. It was found
that for low frequency modes, the spectrum of fluctuations is unaffected
by the higher derivative terms, and the result obtained is the same as
what follows from the
analysis using matching conditions between two Einstein Universes 
\cite{BGGMV,Deruelle:1995kd} joined along a constant scalar field
hypersurface. 

Since the Ekpyrotic scenario makes use of a negative exponential
potential for the scalar matter field, which leads to an extra
instability of the system, it is not clear that the higher derivative 
terms used in \cite{Gasperini:1996fu,Brustein:1997cv,Brustein:1997xn}
can in this case achieve a nonsingular cosmology. The first main
result of this paper is that, with suitably chosen coefficients,
the above mentioned terms are indeed sufficient to produce a
nonsingular cosmology.

In this paper, we add the same higher derivative terms used in 
\cite{Brustein:1997cv} to the action (which includes
a positive or negative potential for the scalar matter field)
and construct nonsingular bouncing cosmologies. At the
level of the effective action, our Lagrangian can be viewed as giving
both nonsingular solutions of modified PBB type (the modification
consisting of the addition of an exponential potential for the
dilaton), and also nonsingular Ekpyrotic solutions. 
The justification for adding these higher derivative terms is different
in the cases of modified PBB cosmology and in the Ekpyrotic
scenario. In the case of PBB cosmology, both the string
coupling constant and the curvature become large as the
dilaton increases, thus justifying the
inclusion of both higher derivative terms of the gravitational action
and of quantum corrections. In the case of Ekpyrotic cosmology (we have
the initial scenario of \cite{Khoury:2001wf} in 
mind in which a bulk brane impacts
our physical space-time orbifold fixed plane at the time of the bounce
and in which the dilaton and hence the string coupling constant are fixed),
the density and hence curvature at the bounce are large, thus justifying
including higher derivative terms. In addition, the brane collision is
a quantum mechanical process, thus justifying including loop corrections
in the action. Note that our method yields a way of constructing
a nonsingular bouncing Universe which works even in a spatially flat
Universe and is thus different from the constructions of 
\cite{Gordon:2002jw} in which a positive spatial curvature is 
used to generate a bouncing cosmology. 
Since tracing back the
spatial curvature into the very early Universe given the present date
leads - under the assumption that there was no period of inflation
after the bounce - to a highly suppressed curvature at early times,
our approach in obtaining a bouncing cosmology appears more realistic.

We follow the fluctuations through the bounce,
and study the spectrum of the resulting cosmological perturbations at
late times. In this analysis, no matching conditions at the
bounce are necessary. Note, however, that in principle the final
spectrum could depend on the frame in which the higher order
correction terms are introduced, and on the specific form of the
correction terms. In our nonsingular scenario discussed in Sec.IV,  
the correction terms are
defined in the string frame and we find
that the final spectrum of cosmological fluctuations on long wavelength
scales has a shape which agrees
with what is obtained when applying the matching conditions of 
\cite{Hwang:1991an,Deruelle:1995kd}
on a constant scalar field surface \footnote{Note that
up to terms of order $k^2$, the constant scalar field
surface and the constant energy density surface are identical, as
discussed in \cite{Brandenberger:2001bs}.} 
(the most physical choice of the
matching surface both in PBB and Ekpyrotic models). In particular, for
the Ekpyrotic model of \cite{Khoury:2001wf} rendered nonsingular by our
construction, we obtain a blue spectrum of the curvature perturbation 
with index $n_{\cal R}= 3$.

\section{Single scalar field with an exponential potential}

The Lagrangian considered in this paper can be used to describe both
a modified PBB model in which the dilaton has an exponential
potential as well as the Ekpyrotic scenario. Our Lagrangian describes
gravity plus a single scalar matter field $\phi$. In the case of
PBB cosmology, the physical frame is the string frame, and 
$\phi$ is the dilaton field. In the case
of the original version of the Ekpyrotic scenario \cite{Khoury:2001wf},
the physical frame is the Einstein frame since the dilaton is fixed, and
the field $\phi$ is related to the separation of a bulk brane from our
four-dimensional space-time orbifold fixed plane. In the case of
the second version of the Ekpyrotic scenario \cite{Khoury:2001bz} and in
the cyclic variant thereof 
\cite{Steinhardt:2001vw,Steinhardt:2001st}, $\phi$ is the modulus
field denoting the size of the orbifold (the separation of the two
orbifold fixed planes). 

We begin with the Lagrangian of the four-dimensional effective 
theory in the string frame, which is:
\be \label{effactions} 
{\cal S}_S \, = \, \int d^4x \sqrt{-g} e^{-\phi} 
\left[ \frac12 R+\frac12 (\nabla \phi)^2-V_S(\phi) \right]\,, 
\ee 
where $R$ is the Ricci scalar, $V_S(\phi)$ is the scalar field potential 
in the string frame. In this form, the action looks reminiscent of the action 
for PBB cosmology. Note that
$V_S(\phi)=0$ in the simplest version of the PBB scenario.  
We set the units such that $8\pi G \equiv 1$ with $G$ being 
a four-dimensional gravitational constant.   
Making a conformal transformation 
\be \label{conformal} 
\hat{g}_{\mu\nu}=e^{-\phi}g_{\mu\nu}\,, 
\ee
the action in the Einstein frame can be written as 
\be \label{effactione}
{\cal S}_E \, = \, \int d^4x \sqrt{-\hat{g}}\left[ \frac12 \hat{R}
- \frac14 (\hat{\nabla} \phi)^2-V_E(\phi) \right]\,, 
\ee
where 
\be \label{ekypo} 
V_E(\phi) \equiv e^{\phi}V_S(\phi) \,. 
 \ee 
Introducing a rescaled field $\varphi=\pm \phi/\sqrt{2}$,
the action (\ref{effactione}) reads
\be \label{effactione2}
{\cal S}_E \, = \, \int d^4x \sqrt{-\hat{g}}\left[ \frac12 \hat{R}
-\frac12 (\hat{\nabla} \varphi)^2-V_E(\phi(\varphi)) \right]\,. 
\ee 
In this form, the action is seen to describe both the PBB
model in the Einstein frame, as well as the Ekpyrotic 
scenario \cite{Durrer:2002jn}.

The Ekpyrotic scenario is characterized by an exponential 
potential \cite{Khoury:2001wf} 
\be \label{einpoten} 
V_E = - V_{0} \exp 
\left(-\sqrt{\frac{2}{p}}\,\varphi\right)\,.  
\ee 
with $0 < p \ll 1$. The field $\varphi$ 
denotes the separation of two parallel branes. 
According to the Ekpyrotic scenario, the branes are 
initially widely separated but are approaching each other, 
which means that $\varphi$ begins near $+\infty$ and is
decreasing toward $\varphi = 0$.  In the PBB scenario, in contrast, the 
dilaton starts out from a weakly coupled regime with $\phi$ increasing
from $-\infty$.  
Thus, if we want the potential (\ref{einpoten}) to describe a
modified PBB scenario with a dilaton potential which is important
when $\phi \to 0$ but negligible for $\phi \to - \infty$, we 
have to use the relation $\varphi=-\phi/\sqrt{2}$ between the field 
$\varphi$ in the Ekpyrotic case and the dilaton $\phi$ in the PBB case.

Adopting the Friedmann-Robertson-Lemaitre-Walker (FRLW) metric 
$ds^2=-dt_E^2+a_E^2dx_E^2$ in the Einstein frame, 
the background equations are given by 
\be \label{basicein} 
3H_E^2=\frac12\dot{\varphi}^2+V_E(\varphi)\,, \\
~~~\ddot{\varphi}+3H_E\dot{\varphi}+V_E'(\varphi)
=0 \,, 
\ee 
where a prime denotes a derivative with respect to a cosmic time,
$t_E$. For the exponential potential (\ref{ekypo}) we have the following exact 
solution 
\be \label{ekysolution} 
a_E \propto (-t_E)^p,~~~H_E=\frac{p}{t_E},~~~ 
V_E=-\frac{p(1-3p)}{t_E^2},~~~ \dot{\varphi}=
\frac{\sqrt{2p}}{t_E} \,.  
\ee 
The solution for $t_E<0$ describes the contracting universe
in the Einstein frame prior to the collision of branes.

In the string frame the action is given by (\ref{effactions})
with potential 
\be \label{ekypos}
V_S= - V_{0} \exp \left[\left(\frac{1}{\sqrt{p}}-1\right) 
\phi\right]\,.  
\ee 
The FRLW metric in the string frame is described by
$ds^2=e^{-\phi}(-dt_S^2+a_S^2dx_S^2)$, which is connected to 
the quantities in the Einstein frame as 
\be \label{ta} 
dt_S=e^{-\varphi/\sqrt{2}}dt_E,~~~ 
a_S=e^{-\varphi/\sqrt{2}}a_E \,, 
\ee 
where we used the relation $\phi=-\sqrt{2}\varphi$.
Integrating the first relation gives
\be 
-(1-\sqrt{p})t_S=(-t_E)^{1-\sqrt{p}} \,. 
 \ee 
Therefore the evolution of $a_S$ and $\phi$ in the string frame is
\be 
\label{ekysolution2}
 a_S \propto (-t_S)^{-\sqrt{p}},~~~ 
\phi=-\frac{2\sqrt{p}}{1-\sqrt{p}}\ln \left[ -(1-\sqrt{p})t_S\right] \,.  
\ee 
This illustrates the super-inflationary solution with growing dilaton 
from $\phi=-\infty$.
Note that singularities are inevitable in both frames as $t \to 0$.
We wish to analyze whether this singularity can be avoided 
by including higher-order corrections.

\section{General actions and evolution equations}

In this section  
we present the background and perturbed equations
in the case of a generalized action containing higher derivative terms. 
We write this action in the form \cite{Cartier:2001is,Hwang:1999gf}
\begin{eqnarray}
 S = \int d^4 x \sqrt{-g} \left[ \frac12 f(R, \phi) - \frac12 \omega 
 (\phi) (\nabla \phi)^2-V(\phi)+{\cal L}_c \right],
\label{lag}
\end{eqnarray}
where $f(R, \phi)$ is a function of the Ricci scalar $R$ and
a scalar field $\phi$.  $\omega(\phi)$ and $V(\phi)$ are general functions
of $\phi$.  The Lagrangian ${\cal L}_c$ represents the higher-order
corrections to the tree-level action. Both higher derivative
gravitational terms and terms involving $\varphi$ appear.
The action (\ref{lag}) applies not only to 
low-energy effective string theories, but also to effective action
approaches to Einstein quantum gravity and to scalar tensor theories, among
others.

As mentioned in the Introduction, our motivation for considering
the addition of higher derivative terms in the effective action is
to construct a nonsingular bouncing model and thus to overcome the 
singularity problem (``graceful exit problem'') which plagues both 
the PBB and the Ekpyrotic scenario. 
The higher-order contribution ${\cal L}_c$ can be written as the sum 
of the $\alpha'$ classical correction ${\cal L}_{\alpha'}$ and the quantum 
loop correction ${\cal L}_{q}$ \cite{Brustein:1997cv,Cartier:2001gc}. 
Both involve the same gravitational and scalar field terms, but are multiplied
by different powers of $e^{-\phi}$.  

The leading $\alpha'$ (string) correction to the
gravitational action we adopt is given 
by \cite{Gasperini:1996fu,Brustein:1997cv}
\begin{eqnarray}
 {\cal L}_{\alpha'} = -\frac12 \alpha' \lambda
  \xi(\phi) \left[ c R_{\rm GB}^2+ d 
 (\nabla \phi)^4 \right]\,,
\label{lagalpha}
\end{eqnarray}
where $\xi(\phi)$ is a general function of $\phi$ and $R_{\rm GB}^2 
=R^2-4R^{\mu\nu}R_{\mu\nu}+ R^{\mu\nu\alpha\beta}R_{\mu\nu\alpha\beta}$ is 
the Gauss-Bonnet term. The inverse string tension, $\alpha'$, is  set to unity.
The Gauss-Bonnet term has the property of keeping the order of the
gravitational equations of motion unchanged. It has been known since
the early days of string theory that this term arises as the lowest
string correction to the gravitational field equations in a string
theory background. At tree level (lowest order in $\hbar$), we have
$\xi(\phi) = -e^{-\phi}$. When applied to PBB cosmology, it turns out
that as $t \to -\infty$, the invariants $R_{\rm GB}^2$ and $(\nabla \phi)^4$
decay faster than the coefficient function $\xi(\phi)$ blows up. Hence,
the correction terms in the Lagrangian are unimportant for large negative 
values of $\phi$, but become important as the system approaches 
the strongly coupled region ($\phi \sim 0$).

Following \cite{Brustein:1997cv}, we take the higher $n$-loop correction
terms ${\cal L}_{q}$ in addition to the
tree-level term ${\cal L}_{\alpha'}$.
For the moment, however, we will keep $\xi(\phi)$ general.
We will give specific forms for $\xi(\phi)$ and ${\cal L}_{q}$ later.

 \subsection{Background equations}

Variation of the action (\ref{lag}) with respect to the scale factor,
the lapse function (then set to $1$ after the variation) and the
scalar matter field leads to 
the following background equations \cite{Cartier:2001is} 
\begin{eqnarray}
& & H^2 = \frac{1}{6F} \left( \omega \dot{\phi}^2+RF-f
+2V-6H\dot{F}+\rho_c \right)\,, \\
& & \dot{H}= \frac{1}{2F} \left(-\omega \dot{\phi}^2+H\dot{F} 
-\ddot{F}-\frac12 \rho_c-\frac12 p_c \right)\,, \\
& & \ddot{\phi}+3H\dot{\phi}+\frac{1}{2\omega}
\left( \omega_{\phi}\dot{\phi}^2-f_{\phi}+2V_{\phi}
-\Delta_{\phi} \right)=0 \,,
\label{back}
\end{eqnarray}
where $F \equiv \partial f/\partial R$ and $H \equiv \dot{a}/a$.
$\rho_c$, $p_c$, and $\Delta_{\phi}$ correspond to the higher-order
curvature and derivative corrections with stress-energy tensor  
$T^{\mu}_{\nu}=(-\rho_c, p_c, p_c, p_c)$. 
$\Delta_{\phi}$ comes from the variation 
of ${\cal L}_c$ with respect to $\phi$.  
For the tree-level $\alpha'$ correction (\ref{lagalpha}), one has 
\begin{eqnarray}
& & \rho_c =2\alpha' \lambda \left( 12c \dot{\xi}H^3
-\frac32 d\xi \dot{\phi}^4 \right)\,,
\label{corre1} \\
& & p_c \equiv -2\alpha' \lambda \left\{ 4c \left[ \ddot{\xi}
H^2+ 2\xi H(\dot{H}+H^2)\right]+\frac12 d\xi \dot{\phi}^4 \right\}\,, \\
& & \Delta_{\phi} \equiv -\alpha' \lambda \left[ 24 c \xi_{\phi}
H^2 (\dot{H}+H^2)-d \dot{\phi}^2(3\dot{\xi}\dot{\phi}
+12\xi \ddot{\phi}+12\xi\dot{\phi}H) \right]\,.
\label{alphacorre}
\end{eqnarray}
Note that taking into account quantum loop corrections provides 
additional source terms for $\rho_c$, $p_c$, and $\Delta_{\phi}$.  
We will discuss this issue in Sec.~IV.

 \subsection{Perturbation equations}

A perturbed space-time
metric has the following form for scalar perturbations 
in an arbitrary gauge (see e.g. \cite{MFB}, where the function
$A$ is denoted by $\phi$):
\begin{eqnarray}
ds^2 = -(1+2A)dt^2 + 2a(t)B_{,i} dx^idt 
+&a^2(t)[(1-2\psi)\delta_{ij}+2E_{,i,j}] dx^i dx^j\,, 
\label{pmetric}
\end{eqnarray}
where a comma denotes the usual flat space 
coordinate derivative. 
We introduce the curvature perturbation, ${\cal R}$, in the comoving gauge
\cite{Lyth:1985} 
\begin{eqnarray}
{\cal R} \equiv \psi+\frac{H}{\dot{\phi}}\delta \phi\,.
\label{metric}
\end{eqnarray}
The perturbed Einstein equations for the action (\ref{lag})
are written in the form \cite{Hwang:1999gf,Cartier:2001is} 
\begin{eqnarray}
\frac{1}{a^3Q} \left(a^3Q \dot{\cal R} \right)^{\bullet} -s 
\frac{\Delta}{a^2} {\cal R}=0\,,
\label{peinstein}
\end{eqnarray}
where
\begin{eqnarray} \label{Qs2}
Q &\equiv& \frac{\omega \dot{\phi}^2 +3I
(\dot{F}-4\lambda c \dot{\xi}H^2)-6\lambda d \xi 
\dot{\phi}^4}{\left( H + I\right)^2}\,,  \\
s &\equiv& 1+\frac{4\lambda c\xi 
\dot{\phi}^4 -16\lambda c \dot{\xi}\dot{H}I+ 8\lambda c 
(\ddot{\xi}-\dot{\xi} H)I^2} {\omega \dot{\phi}^2 
+3I(\dot{F} -4\lambda c \dot{\xi}H^2)
-6\lambda d \xi \dot{\phi}^4}\,.
\label{Qs}
\end{eqnarray}
with $I \equiv (\dot{F}-4\lambda c \dot{\xi}H^2)/ (2F-8\lambda c \dot{\xi} H)$.

Introducing a new quantity, \footnote{Note that $\Psi$ is the variable in
terms of which - for unmodified Einstein gravity - the action for 
fluctuations has the canonical form of a
free field action with time dependent mass (see e.g. \cite{MFB} where
the variable is denoted as $v$).}
$\Psi \equiv z{\cal R}$, with $z \equiv 
a\sqrt{Q}$, each Fourier component of $\Psi$ satisfies the second order 
differential equation 
\begin{eqnarray}
\Psi_k''+\left(sk^2-\frac{z''}{z}\right)\Psi_k=0\,,
\label{Psi}
\end{eqnarray}
where a prime denotes the derivative with respect to conformal time,
$\eta \equiv \int a^{-1}dt$.  In the large scale limit, $|sk^2| \ll |z''/z|$, 
eq.~(\ref{Psi}) is integrated to give 
\begin{eqnarray}
{\cal R}_k =C_k+D_k \int \frac{d\eta}{z^2}\,,
\label{Rk}
\end{eqnarray}
where $C_k$ and $D_k$ are integration constants. 
The curvature perturbation 
is conserved on super-Hubble scales as long as the second term in 
eq.~(\ref{Rk}) is not strongly dominating, as in the case of the single 
field, slow-roll inflationary scenarios.  

If the evolution of $z$ before the bounce is given in the form 
\footnote{Note that as long as the additional terms ${\cal L}_c$ in
the action are negligible, then $\gamma = p/(1 - p)$, and thus
$0 < \gamma \ll 1$ for the collapsing phase of Ekpyrotic cosmology,
and $\gamma = 1/2~(p = 1/3)$ for the collapsing phase of PBB.} 
\begin{eqnarray}
z \propto (-\eta)^{\gamma}\,,
\label{z}
\end{eqnarray}
the second term in eq.~(\ref{Rk})
yields $\int d\eta/z^2 \propto (-\eta)^{1-2\gamma}$.  Therefore curvature
perturbations can be amplified for $\gamma \ge 1/2$ on super-Hubble scales, 
while they are not for $\gamma<1/2$ \cite{Tsujikawa:2001ad}
 (Note that ${\cal R}_k \propto {\rm ln} 
(-\eta)$ for $\gamma=1/2$).  Whether this enhancement occurs or not depends 
on the time evolution of $z$, and therefore on the string
cosmological model.

We need to go to the next order solution of eq.~(\ref{Psi})
in order to obtain the spectrum of curvature perturbations.  If $s$ is 
a positive constant (as it will be in the asymptotic limits), 
the solution for $\Psi_k$ is expressed by the 
combination of the Hankel functions: 
\begin{eqnarray}
\Psi_k=\frac{\sqrt{\pi |\eta|}}{2}\left[ c_1 H_{\nu}^{(1)}
(x) +c_2 H_{\nu}^{(2)}(x) \right]\,,
\label{han}
\end{eqnarray}
where 
\begin{eqnarray}
x \equiv \sqrt{s} k |\eta|\,~~~~\nu \equiv \frac12 |1-2\gamma|\,.
\label{def}
\end{eqnarray}
The solution (\ref{han}) corresponds to the Minkowski 
vacuum state in the small-scale limit ($k \to \infty$).

We can expand the Hankel functions in the following 
form \cite{Hwang:2002ks}:
\begin{eqnarray} 
H_{\nu}^{(1, 2)} (x)=\sum_{n=0}^\infty \frac{1}{n!}
\left(-\frac{x^2}{4}\right)^n \frac{1}{\sin \pi \nu}
\left[ \left(\frac{x}{2}\right)^{\nu} \frac{\pm ie^{\mp i
\pi \nu}}{\Gamma(\nu+n+1)}+\left(\frac{x}{2}\right)^{-\nu} \frac{\mp 
i}{\Gamma(-\nu+n+1)}\right]\,.
\label{Hankel}
\end{eqnarray}
The curvature perturbation, ${\cal R}_k=
\Psi_k/z$, has two solutions which are proportional to 
$k^{\nu}|\eta|^{\nu-\gamma+1/2}$
and $k^{-\nu}|\eta|^{-\nu-\gamma+1/2}$, which follow from the 
first and second term in eq.~(\ref{Hankel}), respectively.
In the large-scale limit ($k \to 0$), the contribution of 
the second term dominates over the first one, 
thereby yielding the spectrum of the curvature perturbation as 
\begin{eqnarray}
P_{{\cal R}} \equiv \frac{k^3}{2\pi^2}\left| 
{\cal R}_k \right|^2 \propto k^{3-2\nu} \propto k^{n_{\cal{R}}-1}\,,
\label{PR}
\end{eqnarray}
in which case the spectral tilt is 
\begin{eqnarray}
n_{\cal{R}}-1=3-\left| 1-2\gamma \right|\,.
\label{ind}
\end{eqnarray}
Note that we have $k^{-\nu}|\eta|^{-\nu-\gamma+1/2}=k^{-\nu}|\eta|^0$
for $\gamma<1/2$, in which case the constant mode $C_k$ in 
eq.~(\ref{Rk}) corresponds 
to the solution which comes from the second term in eq.~(\ref{Hankel}).
The Ekpyrotic scenario with a negative potential 
($0<p<1/3$) belongs to this case ($\gamma<1/2$)
as we will show later. 
When $\gamma>1/2$ one has $k^{-\nu}|\eta|^{-\nu-\gamma+1/2}=
k^{-\nu}|\eta|^{1-2\gamma}$, which means that ${\cal R}_k$ grows 
before the graceful exit ($\eta<0$). We will discuss a 
string-inspired model that belongs to this case in Sec.~V.
The PBB scenario corresponds to the marginal case with 
$\gamma=1/2$.

Note that $s$ is exactly unity in eq.~(\ref{Psi}) when
the corrections ${\cal L}_c$ are not taken into account. 
In the presence of higher-order corrections (${\cal L}_c \ne 0$),  $s$ is 
generally a time-varying function, in which case the formula (\ref{ind}) can 
not be directly applied. 
Nevertheless it is still valid if $s$ is a slowly varying positive function.

In subsequent sections we shall apply the above general formulas 
to concrete string-inspired models. In Section IV we apply the
string loop and quantum corrections to the low energy effective action
in the string frame and find nonsingular bouncing cosmological solutions.
In Section V, we consider a situation with fixed dilaton and add a
Gauss-Bonnet term to the Einstein frame action. We find nonsingular
cosmological solutions which begin in an asymptotically flat state,
undergo a period of super-exponential inflation which terminates
with a graceful exit.

\section{Inclusion of higher-order corrections in the string 
frame \\
--dilaton-driven case} 

In the context of PBB cosmology, the natural frame to use
in order to define the correction term ${\cal L}_c$ in the
Lagrangian is the string frame. In this case, we should use
$f=e^{-\phi}R$ and $\omega=-e^{-\phi}$, i.e. 
\be \label{actionloop} 
{\cal S}_S \, = \, \int d^4x \sqrt{-g}\left\{ e^{-\phi} \left[ \frac12 
R+\frac12 (\nabla \phi)^2-V_S(\phi) \right]+{\cal L}_c\right\} \,,
\ee 
where $\phi$ corresponds to the dilaton. This Lagrangian was
suggested in \cite{Gasperini:1996fu,Brustein:1997cv}, and used
to construct nonsingular background cosmological solutions of
PBB cosmology in the absence of a potential for the dilaton.
Fluctuations in this model were studied 
in \cite{Cartier:2001is}
in the absence of a dilaton potential. 

In the following, we
extend these analyses to the case of a non-vanishing dilaton
potential. We will first construct nonsingular solutions which
in the Einstein frame correspond to nonsingular bouncing cosmologies.
We then study how the fluctuations evolve across the bounce and
compute the spectrum of fluctuations. The analysis in this section
thus applies immediately to the modified PBB scenario in which the
dilaton has a negative exponential potential. We can also apply
the results to the initial version \cite{Khoury:2001wf}
of the Ekpyrotic scenario. In
this case, the brane collision occurs at $\varphi = 0$, and since
thus the gravitational coupling constant does not change significantly
near the bounce, the difference in the role of the higher derivative
terms between the Einstein frame (the frame in which it appears
most logical to define the correction terms in the Lagrangian) and
the ``string frame'' (quotation marks used here because in the case
of the Ekpyrotic scenario $\varphi$ is not the dilaton, the dilaton
being fixed) is not expected to be significant. The application of
the results of this section to the version of the Ekpyrotic scenario
with moving boundary branes \cite{Khoury:2001bz} and to the cyclic
scenario \cite{Steinhardt:2001vw} is more problematic since in
this case $\varphi$ is the dilaton, $\varphi \to -\infty$ at the
bounce, and thus the difference in the evolution of models with
correction terms defined in the Einstein and string frames is
expected to be important. 

The reason why nonsingular solutions are possible in
the presence of the correction term ${\cal L}_c$, is that such a term
can lead to violations of the null energy condition (from the
perspective of an observer using unmodified Einstein equations). 
Thus, it is expected to lead to a successful graceful exit, in
the same way that introducing matter violating the null energy
condition allowed the construction of nonsingular bouncing models in
\cite{Hwang:2001zt,Peter:2002cn}. 

In this model the background equations 
are written as 
\beqa \label{motionse} 
6H^2-6H\dot{\phi}+\dot{\phi}^2-2V_S=e^{\phi}\rho_c 
\,, \\
4\dot{\phi}H-4\dot{H}-6H^2-\dot{\phi}^2+2\ddot{\phi}+2V_S=
e^{\phi}p_c \,, \\
6\dot{H}+12H^2+\dot{\phi}^2-2\ddot{\phi}-6H\dot{\phi}
-2(V_S-V_S')=e^{\phi} \Delta_{\phi}\,. 
\label{motionse2} 
\eeqa 
The dilatonic corrections ${\cal L}_c$ are the sum of the tree-level 
$\alpha'$ corrections and the quantum $n$-loop corrections ($n=1, 2, 3, 
\cdots$), with the function $\xi(\phi)$ [see (\ref{lagalpha})] given by
\beqa 
\xi(\phi)=-\sum_{n=0} C_n e^{(n-1)\phi} \,,
\label{xifunction} 
\eeqa 
where $C_n$ ($n \ge 1$) are the coefficients of 
$n$-loop corrections with $C_0=1$.
In this case the source terms due to ${\cal L}_c$ on the right hand side of 
(\ref{motionse})-(\ref{motionse2}) are given by \cite{Cartier:2001gc}
\begin{eqnarray} \label{C_n}
\rho_c=\sum_{n=0} C_n \left\{\rho_c \right\}_n\,,~~~
p_c=\sum_{n=0} C_n 
\left\{p_c \right\}_n,~~~ \Delta_\phi=\sum_{n=0} C_n 
\left\{\Delta_\phi \right\}_n\,,
\label{sum}
\end{eqnarray}
where 
\begin{eqnarray}
\left\{ \rho_c \right\}_n &=& \alpha' \lambda
\dot{\phi} e^{(n-1)\phi} 
\left\{-24c(n-1)H^3+3d\dot{\phi}^3 \right\}\,, \\
\left\{p_c \right\}_n &=& \alpha' \lambda e^{(n-1)\phi} \left\{8c(n-1)H 
\left[(n-1)\dot{\phi}^2H+\ddot{\phi}H+ 2\dot{\phi}(\dot{H}+H^2) \right] 
+d\dot{\phi}^4 \right\}\,, \\
\left\{\Delta_{\phi} \right\}_n &=& \alpha' \lambda  e^{(n-1)\phi} 
\left\{24c(n-1)H^2(\dot{H}+H^2) -3d\dot{\phi}^2 
\left[4\ddot{\phi}+4\dot{\phi}H+ (n-1)\dot{\phi}^2 \right] \right\}\,,
\label{corre}
\end{eqnarray}
with $\lambda=-1/4$.  Following ref.~\cite{Cartier:2001is} 
we choose the coefficients 
as $c=-d=-1$.  Note that the above corrections include the $\alpha'$ 
corrections (\ref{corre1})-(\ref{alphacorre}), corresponding to $n=0$.  

It is also convenient to relate the Hubble parameter and its derivative in the 
Einstein frame with those in the string frame by using eq.~(\ref{ta})
\begin{eqnarray} 
H_E = e^{\phi/2}\left(H_S-\frac{\dot{\phi}}{2}\right),~~~~ 
\dot{H}_E = e^{\phi} \left( \dot{H}_S-\frac{\ddot{\phi}}{2}
+\frac12 \dot{\phi}H_S-\frac{\dot{\phi}^4}{4}\right)\,.
\label{relation}
\end{eqnarray}
Here the dots on the right hand side 
denote the time-derivative with respect to $t_S$. 
The energy density $\rho_E$ and the pressure $p_E$ in the Einstein frame are 
expressed as 
\begin{eqnarray}
\rho_E=3H_E^2\,,~~~p_E=-3H_E^2-2\dot{H}_E\,.
\label{relation2}
\end{eqnarray}
Once we know the evolution of the background in the string frame, 
it is easy to find the evolution of $H_E, a_E, \varphi=-\phi/\sqrt{2}$
and to check whether the null energy condition, $\rho_E+p_E>0$,
holds or not in the Einstein frame by using eqs.~(\ref{relation})
and (\ref{relation2}).  Note that in the absence of higher-order corrections
(${\cal L}_c=0$) one has $2\dot{H}_E=-(\rho_E+p_E)=-\dot{\phi}^2<0$.  
In this case once the contraction begins ($\dot{H}_E<0$) the Hubble 
parameter is {\it always} negative.  Therefore it is not possible to 
have bouncing solutions
required for the nonsingular Ekpyrotic scenario unless higher-order 
corrections ${\cal L}_c$ are taken into account.

 \subsection{Background evolution} 
 
In the absence of a negative exponential potential ($V_S=0$),
it was found in Ref.~\cite{Gasperini:1996fu} that curvature singularities 
can be avoided by taking into account higher-order corrections ${\cal L}_c$.
In this case we have nonsingular bouncing solutions
in the Einstein frame due to the violation of the null energy condition. 
We are interested in whether singularity avoidance is possible or not 
in the presence of the Ekpyrotic potential (\ref{ekypos}). Note that
since near the bounce $H^2 \sim t^{-2}$, higher curvature corrections
to the Einstein action will likewise be important in the presence of a
potential.

When $V_S \ne 0$ and ${\cal L}_c=0$ the background solutions are described 
by eqs.~(\ref{ekysolution}) and (\ref{ekysolution2}).  In the string frame 
the scale factor evolves super-inflationary with growing Hubble rate 
($\dot{H}_S>0$).  We plot in Fig.~\ref{dilaton1} the evolution of 
background quantities both in the string and Einstein frames [see the case 
(i)].  The dilaton $\phi$ starts out from the weakly coupled regime 
$g^2_{string}\equiv e^{\phi} \ll 1$, corresponding to widely separated 
branes in the Ekpyrotic scenario, $\varphi=-\phi/\sqrt{2} \gg 1$.  
In the Einstein frame the Universe 
is contracting with a negative Hubble rate.  The solution inevitably 
meets a curvature singularity as $\phi$ grows toward the strongly coupled 
regime ($g^2_{string} \sim 1$).

Our first main finding is that with $V_S(\phi) \ne 0$ there exist nonsingular 
trajectories  in the presence of higher-order 
corrections (${\cal L}_c \ne 0$).  
Thus, the presence of the potential for the dilaton
does not prevent the higher derivative terms from being able to
smooth out the curvature singularity. The details
depend on the value of the power-law index, $p$.
When $p \ll 1$ the Ekpyrotic potential (\ref{einpoten}) is exponentially 
suppressed for $\varphi~\gsim~1$, in which case the dynamics of 
the system is hardly affected by the negative potential except for 
the region, $\varphi \sim 0$. However, in this region the higher
derivative terms play a crucial role. 

In our simulations, we have adopted the potential 
(\ref{einpoten}) for $\varphi>0$ and $V_E=0$ for $\varphi<0$. This
is in the spirit of the first version of the Ekpyrotic scenario
\cite{Khoury:2001wf} in which the potential vanishes at the brane 
collision, the bulk brane is absorbed by the orbifold fixed
plane via a small instanton transition, and there is no potential
left afterwards.   We show in Fig.~\ref{dilaton1} the dynamical 
evolution of the system for $p=0.1$. The case (i) is the one in
which only tree-level terms are present and in which singularity
avoidance is not possible.  
The case (ii) corresponds to the one where both tree-level and one-loop 
corrections are taken into account ($C_1=1.0$ and $C_2=0$).  Inclusion of 
one-loop corrections makes it possible to have nonsingular cosmological 
solutions.  In fact $\rho_E+p_E$ becomes negative around $t_S \sim 115$ in 
Fig.~\ref{dilaton1}, after which the Hubble parameter $H_E$ begins to grow.  
The Universe starts to expand once $H_E$ crosses zero.  Namely the 
violation of the null energy condition allows to have nonsingular bouncing solutions in the 
Einstein frame.  Nevertheless we should notice that the scale factors 
evolve super-inflationary both in the string and Einstein frames due to 
unbounded increase of $H_S$ and $H_E$, together with rapid growth of the 
field $\phi$.  Therefore we are faced with another problem, namely how 
to connect to the stage of a decreasing Hubble parameter.

If two loop terms are added
(keeping the previous tree-level and one loop terms)
phenomenologically more appealing nonsingular solutions can be
obtained.  When  $C_2$ is positive, the evolution of 
the system does not differ significantly compared to the case (ii).
However, it is possible to obtain a decreasing
Hubble rate if we take a negative value of $C_2$.
The case (iii) of Fig.~\ref{dilaton1} corresponds to the 
coefficients $C_1=1.0$ and $C_2=-1.0 \times 10^{-3}$.  
We find that the growth rates of the scale factor and of $\phi$ are slowed 
compared to the case (ii). 
We see that $\rho_E+p_E$ becomes positive and begins to 
decrease toward $+0$ after the short period of violations of 
the null energy condition.
Although this case does not correspond to the radiation-dominated Universe
after the graceful exit, it is possible to connect to it by taking into 
account the decay of the dilaton to radiation \footnote{If we were
to include production of radiation at a fixed time during the expanding phase,
we could use the well-known results on the constancy of ${\cal R}$ in
the expanding phase \cite{Lyth:1985,Brandenberger:1984,Bardeen:1983} to  
argue that the spectrum of fluctuations on large scales will be the same
as what is obtained in this paper. The crucial fact about our bounce is
that it is not symmetric in time (see Fig.~1).}. However, including  
radiation in the Ekpyrotic cosmology has some subtle points,
and we do not consider this problem in the present work. 

We have checked that the addition of three loop terms with coefficients
chosen to be of the order $10^{-7}$ (roughly the same hierarchy of
coefficients between the two and three loop terms as between the
one and two loop terms)  does not change the results of the two loop
analysis in a significant way. With a coefficient of the three loop
term of the order 1, the background solution ceases to be nonsingular.

We emphasize that we have nonsingular bouncing solutions 
in the Einstein frame even in the presence of a negative exponential
potential.  When $p \ll 1$ the potential is vanishingly small for 
$\varphi \gg 1$, in which case the dynamics of the system is practically 
the same as that of the zero potential discussed in 
Ref.~\cite{Brustein:1997cv}.
In this case the dilaton starts out from the low-curvature regime $|\phi| 
\gg 1$, which  is followed by the string phase with linearly growing dilaton 
and nearly constant Hubble parameter.
During the string phase one has \cite{Gasperini:1996fu} 
\begin{eqnarray} 
a_S \propto (-\eta_S)^{-1},~~~~
\phi=-\frac{\dot{\phi}_f}{H_f} {\rm ln} (-\eta_S)
+{\rm const}\,,
\label{const}
\end{eqnarray}
where $\dot{\phi}_f \simeq 1.40$ and $H_f \simeq 0.62$.
In the Einstein frame this corresponds to a contracting Universe with
\begin{eqnarray} 
a_E \propto (-\eta_E)^{\dot{\phi}_f/(2H_f)-1}\,.
\label{scaleein}
\end{eqnarray}

On the other hand, we can consider the scenario where the negative Ekpyrotic 
potential dominates initially but the higher-order correction becomes 
important when two branes approach sufficiently.  Numerically we confirmed 
that it is possible to have nonsingular solutions 
(see Fig.~\ref{dilaton2}).  
In the simulations we included the correction terms 
of ${\cal L}_c$ only for $\varphi~\lsim~1$.  In this case the background 
solutions are described by eq.~(\ref{ekysolution}) or (\ref{ekysolution2}) 
before the higher-order correction terms begin to work.  Given this 
background solution, one can obtain the spectra of curvature perturbations 
analytically, as we will see in the next section.  The spectra depend on 
whether the higher-order terms are always dominant or not relative to the 
negative potential before the bounce.

\begin{figure}
\begin{center}
\singlefig{14cm}{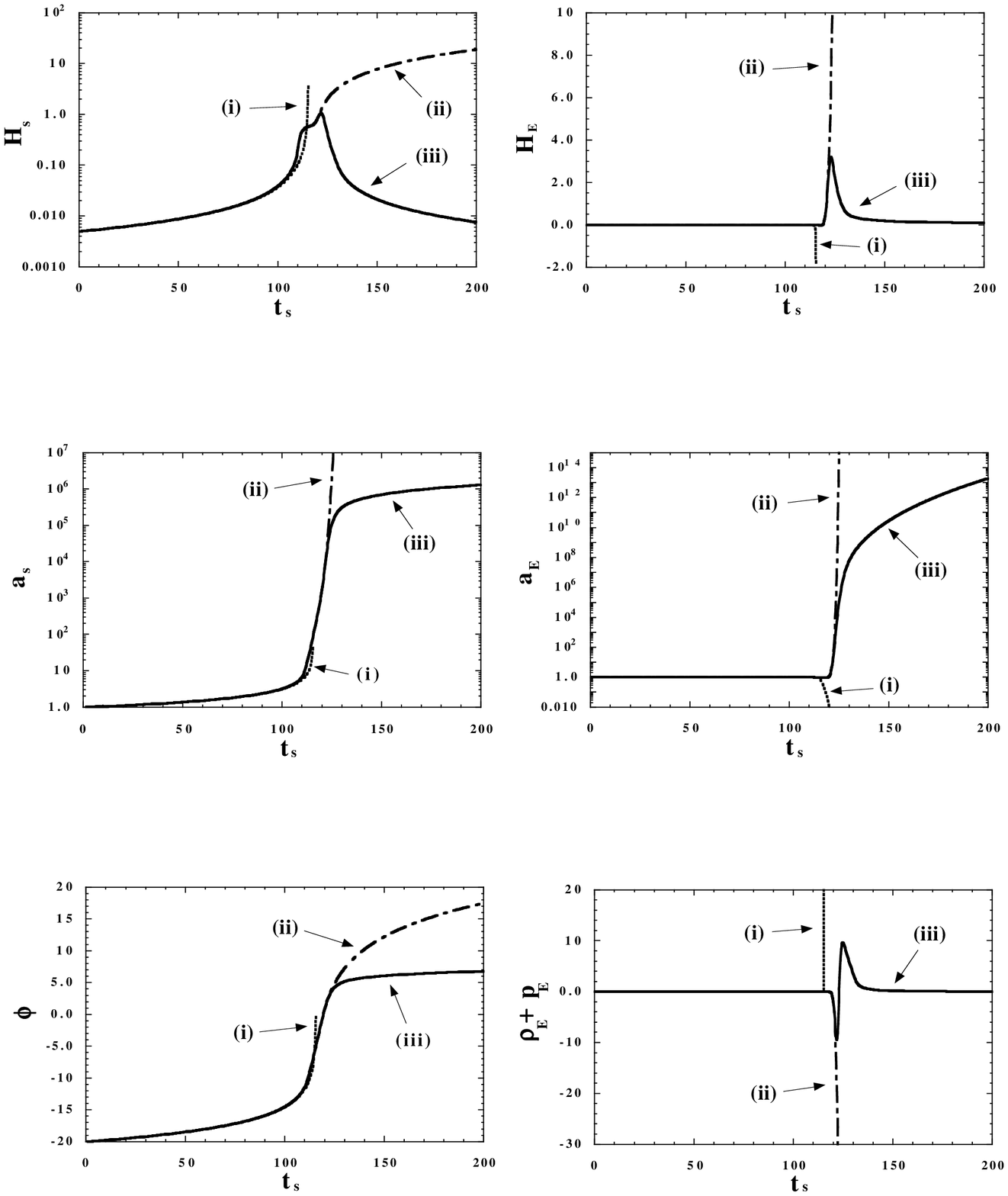}
\begin{figcaption}{dilaton1}{14cm}
The evolution of $H_S$, $H_E$, $a_S$, $a_E$, $\phi$, and
$\rho_E+p_E$ with $c=-1$, $d=1$, $p=0.1$.
We choose initial conditions $\phi=-20$, $H=5.0 \times 10^{-3}$.
The cases correspond to (i) only
tree-level correction terms but no higher-order 
corrections ($C_1=C_2=0$), (ii) tree-level and one-loop corrections 
present ($C_1=1.0, C_2=0$) (iii) tree-level and one- and two-loop 
corrections present with $C_1=1.0$ and $C_2= -1.0 \times 10^{-3}$.
\end{figcaption}
\end{center}
\end{figure}
\begin{figure}
\begin{center}
\singlefig{9cm}{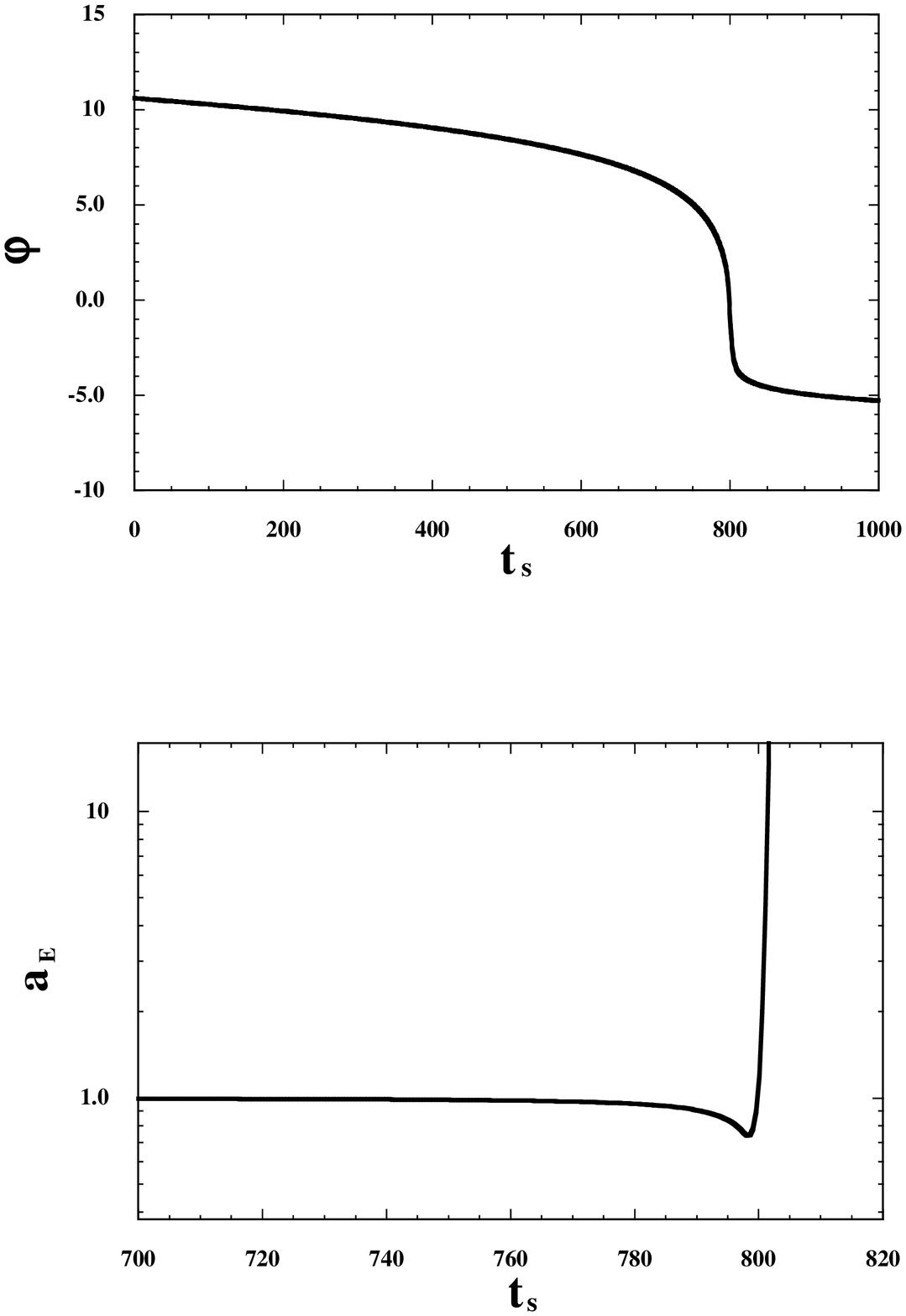}
\begin{figcaption}{dilaton2}{9cm}
The evolution of $\varphi$ and $a_E$  
with $c=-1$, $d=1$, $p=0.1$, $C_1=1.0$ 
and $C_2= -1.0 \times 10^{-3}$.  In this 
case we include the correction term ${\cal L}_c$ only for $\varphi<1$.  
We choose initial conditions $\phi=-15$, $H=1.5 \times 10^{-3}$.
Prior to the collision of branes at $\varphi=0$, the universe is slowly 
contracting, which is followed by the bouncing solution 
through higher-order corrections.
\end{figcaption}
\end{center}
\end{figure}

We have also studied Ekpyrotic potentials with other values of $p$, and
found that if the potential is negative, corresponding to $p<1/3$, then 
singularity avoidance is possible for suitable choices of $C_1$ and $C_2$ 
as in the case $p = 0.1$ shown in Fig.~\ref{dilaton1}.   
In the case of $p>1/3$ the field $\phi$ climbs up a positive
exponential potential due to the Hubble contraction term. 	 
When $p~\gsim~1/2$ with $V_0=p(1-3p)$ we found that the field $\phi$ returns 
back before it reaches the strongly coupled region, $\phi~\gsim~0$.  
This is equivalent to the fact that two parallel branes do not approach 
each other sufficiently.  In such cases the positive exponential potential
makes the field bounce back before the higher-order correction 
becomes important (this may be related to the instability discussed
recently in \cite{Heard:2002dr}).  
If we choose smaller values of $V_0$, it is possible to 
have nonsingular bouncing solutions which are similar to 
Fig.~\ref{dilaton1}.  This case corresponds to the one where the effect of 
the positive potential is negligible compared to higher-order corrections, 
in which case the background solutions are given by eq.~(\ref{const}).  When 
the positive potential is dominant from the beginning, it is difficult to 
obtain a solution where a successful graceful exit is realized by 
higher-order corrections.

 \subsection{Density perturbations}

Let us proceed to the analysis of the evolution and 
the spectra of density perturbations.
We shall consider two cases: (i) the effect of the potential $V(\phi)$ is 
always negligible relative to the correction term ${\cal L}_c$, and (ii) 
the effect of the correction term ${\cal L}_c$ becomes important only 
around the graceful exit ($\phi \sim 0$). Note that the second case
is the physically more interesting one for applications to Ekpyrotic
cosmology.

 \subsubsection{{\rm Case (i)}: $|V(\phi)| \ll |{\cal L}_c|$}

When the correction terms (\ref{corre1})-(\ref{alphacorre})
always dominate relative to the exponential potential (\ref{einpoten}),
the spectra of density perturbations are similar to the ones discussed in 
ref.~\cite{Cartier:2001is}.  During the string phase with linearly growing 
dilaton and nearly constant Hubble parameter with $\dot{\phi}_f \simeq 
1.40$ and $H_f \simeq 0.62$, we have a sufficient amount of inflation with 
e-folds $N \equiv {\rm ln} (a/a_i) >60$ provided that the dilaton field 
satisfies $|\phi| \gg 1$ initially \cite{Cartier:2001is,Cartier:2001gc}.  
In this stage $Q$ defined in eq.~(\ref{Qs}) is proportional to $e^{-\phi}$ 
by making use of eq.~(\ref{const}), thereby leading to 
\begin{eqnarray} 
z \propto (-\eta_S)^{\gamma}\,,~~~{\rm with}~~~
\gamma=-1+\frac{\dot{\phi}_f}{2H_f}\simeq 0.13\,.
\label{gamma}
\end{eqnarray}
Making use of the relation (\ref{ind}) which is valid for positive $s$, the 
spectral tilt of the large-scale curvature perturbation is 
\begin{eqnarray} 
n_{\cal{R}}-1=3-\left| 3- \frac{\dot{\phi}_f}{H_f} \right| 
\simeq 2.26\,.
\label{tilt1}
\end{eqnarray}

The evolution of the frequency shift $s$ is nontrivial (see Fig.~\ref{s}). 
In the low curvature regime where the higher-order terms are not
important, $s$ is positive ($s \simeq 1$), as in the usual
PBB scenario.  It then changes sign and becomes 
negative during a short transition from the low-curvature regime to the 
string phase.  During the string phase, $\dot{\phi}$ and 
$H_S$ are constant ($\dot{\phi} \simeq 1.40$ and $H_S \simeq 0.62$), 
and $\xi \sim -e^{-\phi}$. The correction term on the
right hand side of eq.~(\ref{Qs}) for $s$ dominates
in this phase.
It follows from eq.~(\ref{Qs}) that the $\phi$ dependence of the 
leading term cancels out between $\xi(\phi)$ and $\omega(\phi)$, 
and that hence $s$ is
constant and negative until the graceful 
exit.  In a stage with negative constant $s$, the solution of 
eq.~(\ref{Psi}) can be written in the form 
\begin{eqnarray}
\Psi_k=\sqrt{|\eta|} 
\left[c_1 I_{\nu} (x) +c_2 K_{\nu} (x) \right]\,,
\label{Psid}
\end{eqnarray}
where $x$ and $\nu$ are given in (\ref{han}) (with $s$ replaced by the
absolute value of $s$), and $I_{\nu}$ and $K_{\nu}$ are modified 
Bessel functions, whose asymptotic solutions are $I_{\nu} \propto x^{\nu}$, 
$K_{\nu} \propto x^{-\nu}$ for $x \to 0$, and $I_{\nu} \sim e^x/\sqrt{2\pi 
x}$, $K_{\nu} \sim \sqrt{\pi/(2x)}e^{-x}$ for $x \to \infty$.  
Then one reproduces the spectral tilt (\ref{tilt1}) in the large-scale limit 
($|sk^2| \ll |z''/z|$).  For small-scale modes curvature perturbations show 
exponential instability due to negative frequency shift.  After the horizon 
crossing ($|sk^2|~\lsim~|z''/z|$), curvature perturbations are frozen, 
since $\gamma$ is smaller than $1/2$ in this case.

It was shown in ref.~\cite{Cartier:1999vk} that the ratio 
$\dot{\phi}_f/H_f$ is required to lie in the range 
$2 \le \dot{\phi}_f/H_f \le 3$ for the successful graceful exit in the presence 
of other forms of higher-order $\alpha'$ correction.  Therefore the 
spectral tilt lies in the range 
\begin{eqnarray} 
2 \le n_{\cal{R}}-1 \le 3\,,
\label{tilt2}
\end{eqnarray}
which is valid for large-scale modes ($|sk^2| \ll |z''/z|$).
Therefore we have blue-tilted spectra as long as the correction ${\cal L}_c$
dominates compared to the exponential potential.

\begin{figure}
\begin{center}
\singlefig{10cm}{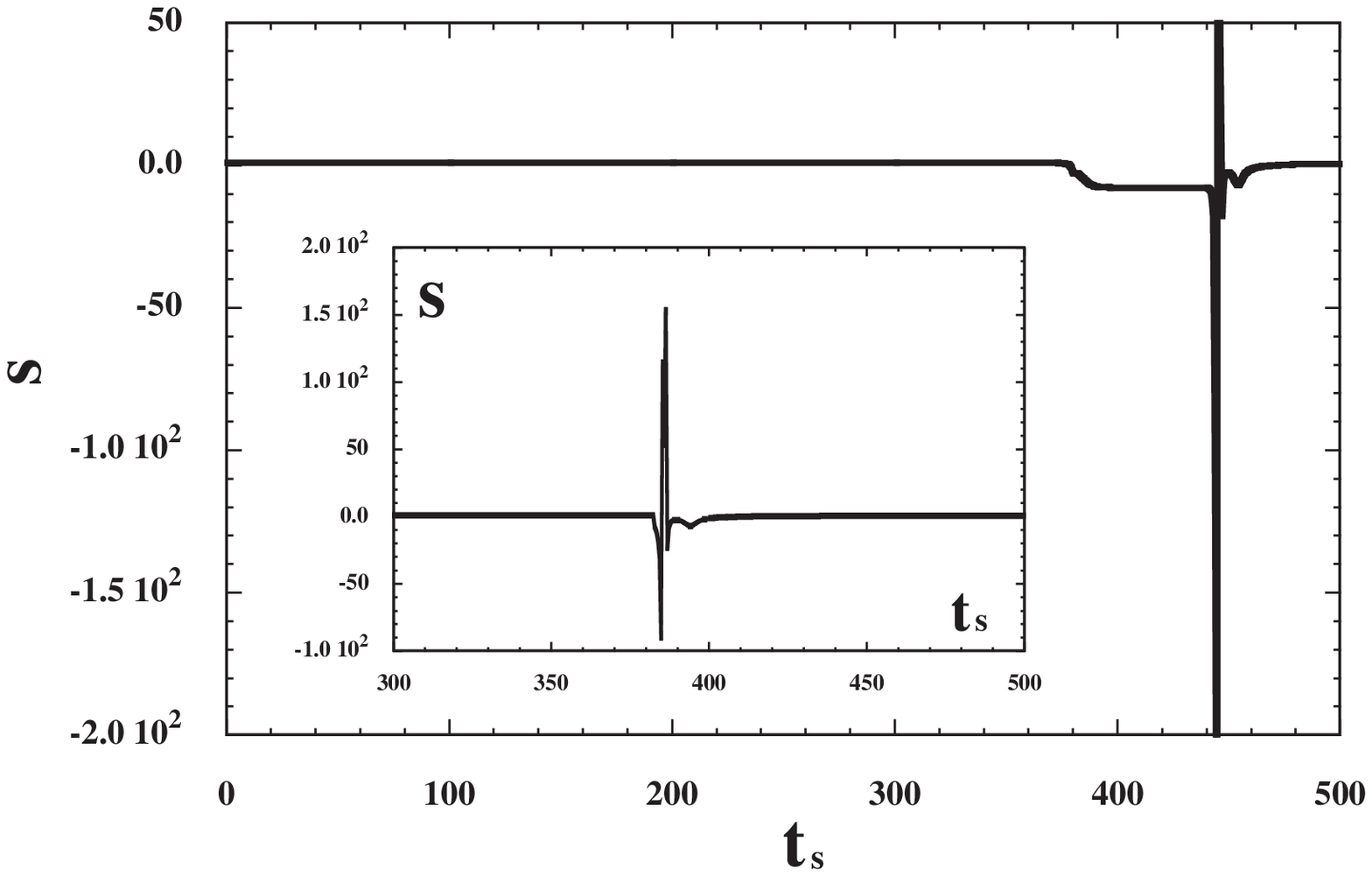}
\begin{figcaption}{s}{10cm}
The evolution of the frequency shift $s$
for $c=-1$, $d=1$, $p=0.1$ with initial 
conditions $\phi=-100$, $H=1.5 \times 10^{-3}$.  
We include the quantum correction ${\cal L}_c$ from the beginning.  
The shift $s$ is approximately constant (and negative) 
during the string phase, which is followed by the 
stage of decreasing curvature with positive $s$.  {\bf Inset}: The 
evolution of $s$ in the case where the quantum correction is taken into 
account only for $\varphi~\lsim~1$.  Note that $s$ rapidly changes sign 
around the graceful exit.
\end{figcaption}
\end{center}
\end{figure}

 \subsubsection{{\rm Case (ii)} : $|V(\phi)| \gg |{\cal L}_c|$
 {\rm but} $|V(\phi)|<|{\cal L}_c|$ {\rm for} $\varphi \sim 0$}

When the correction term ${\cal L}_c$ becomes important only around the 
graceful exit ($\phi \sim 0$), the spectra of density perturbations 
generated before the bounce are mainly determined by the exponential 
potential.  In this case the evolution of the background can be 
characterized by eq.~(\ref{ekysolution2}).  Then the quantity $Q$ in 
eq.~(\ref{Qs2}) evolves as 
\begin{eqnarray}
Q=\frac{2\dot{\phi}^2e^{-\phi}}{(2H_S-\dot{\phi})^2}
 \propto (-\eta_S )^{2\sqrt{p}/(1-p)}\,.
\label{Qstring}
\end{eqnarray}
Therefore we find  
\begin{eqnarray} 
z \propto (-\eta_S)^{\gamma}\,,~~~{\rm with}~~~
\gamma=\frac{p}{1-p}\,,
\label{gamma_po}
\end{eqnarray}
and the spectral tilt for the curvature perturbation is 
\begin{eqnarray} 
n_{\cal{R}}-1= \cases{
         \frac{2}{1-p} & (for $0<p<1/3$)\,, \cr
          \frac{4-6p}{1-p} & (for $1/3<p<1$)\,. \cr
          }
\label{tilt_po}
\end{eqnarray}
This coincides with the result in the Einstein frame
performed in 
ref.~\cite{Lyth:2001pf,Brandenberger:2001bs,Hwang:2001ga,Tsujikawa:2001ad}.
For very slow contraction with a negative Ekpyrotic potential
($p \ll 1$), one has blue tilted spectra with $n_{\cal{R}}-1=2$.  
Since $\gamma$ is less than $1/2$ for $p<1/3$ (i.e., negative potential),
curvature perturbations are not enhanced in the large-scale limit even in 
the presence of the correction ${\cal L}_c$ around the graceful exit.  
The simplest PBB scenario with zero potential corresponds to 
$p=1/3$ and $\gamma=1/2$, in which case one has $n_{\cal{R}}-1=3$.
In this case curvature perturbations evolve as ${\cal R}_k \propto
{\rm ln}(-\eta)$ as found from eq.~(\ref{Rk}) with eq.~(\ref{z}).

We have solved the evolution equation (\ref{peinstein}) for 
the cosmological fluctuations numerically. 
Experience from studying fluctuations
in inflationary cosmology teaches us that following the evolution
equation for $\Psi$ instead of for the gravitational potential $\Phi$
is less likely to be effected by numerical noise.  Since in a
contracting Universe, one of the
modes of $\Phi$ increases much more rapidly than the dominant mode of
$\Psi$, we believe that it is advantageous to use $\Psi$ in our
case as well.  In addition, from a more conceptual point of view, the variable 
${\cal R}$ is preferable since it is more closely related to $\Psi$ 
in terms of which the action for cosmological fluctuations takes on its 
canonical form. Note that $\Psi$ is also the good variable to use when 
following cosmological fluctuation from inflation through reheating
\cite{Finelli:1998bu}.
In Fig.~\ref{speevolution} we plot the resulting evolution of the spectra of 
curvature perturbations for several different frequencies.  The 
higher-order correction ${\cal L}_c$ is included only when two branes 
approach sufficiently, i.e., $\varphi~\lsim~1$.  We find that large-scale 
modes ($k \ll 1$) are not enhanced as predicted by  eq.~(\ref{Rk}).
In contrast, small-scale 
curvature perturbations exhibit rapid growth around the graceful exit 
($\varphi \sim 0$).

There are two reasons for this instability. 
The first is the fact that the frequency shift $s$ 
becomes negative for a short period where the higher-curvature effect is 
dominant (see the inset of Fig.~\ref{s}). As is obvious from 
(\ref{Psi}),  an exponential instability
for $\Psi$ is induced by negative $s$, which is stronger for larger $k$.  
We expect this instability will become important for modes with $\vert s 
\vert_{\rm max}\,k^2~\gsim~1$, where $\vert s \vert_{\rm max}$ is the maximal 
negative value of the function $s$.  {}From the inset of 
Fig.~\ref{s} the maximal absolute value of $s$ during the negative 
branch is about $\vert s \vert_{\rm max} \sim 10^2$.  Hence, the 
$s$-instability is expected to be important only for modes with 
$k~\gsim~10^{-1}$.  By comparing runs with $s$ given by the general 
formula and 
runs with $s = 1$, we were able to determine numerically that the actual 
cutoff value of $k$ below which the instability due to the $s$-term is 
negligible is $k \sim 10^{-2}$.  Thus, we conclude that the main source of 
the short wavelength instability of the fluctuation modes around the bounce 
must be a second one, namely the nontrivial nature of the bouncing 
background and its result on the quantity $z''/z$.
 
After the transition to the 
expanding Universe, the curvature perturbation is nearly
conserved as found in Fig.~\ref{speevolution}.

\begin{figure}
\begin{center}
\singlefig{11cm}{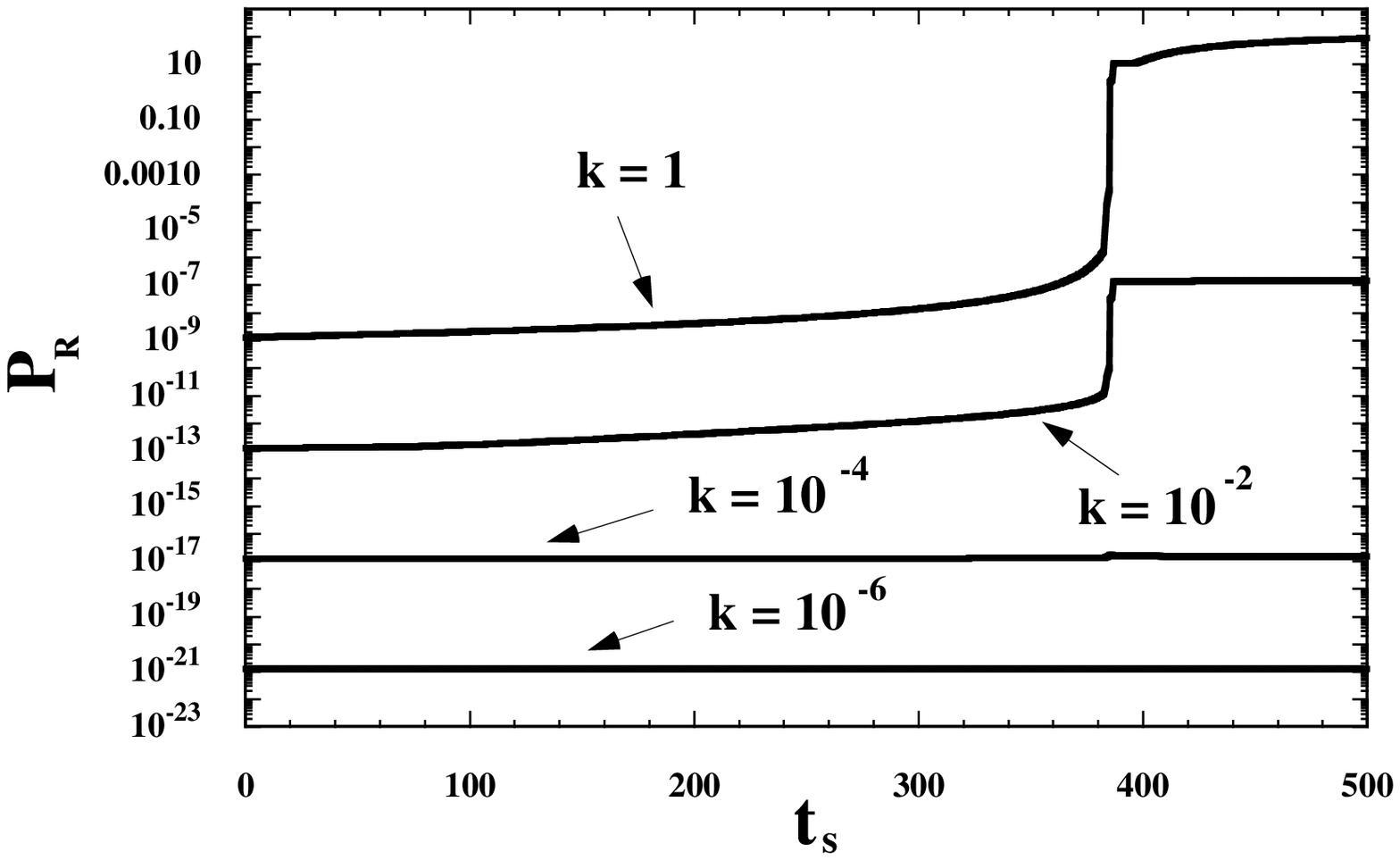}
\begin{figcaption}{speevolution}{11cm}
The evolution of the spectra of curvature perturbations, $P_{\cal R}$,
for $c=-1$, $d=1$, $p=0.1$, $C_1=1.0$ and $C_2= -1.0 
\times 10^{-3}$.  The initial conditions are chosen to be $\phi=-15$, 
$H=1.5 \times 10^{-3}$.  We include the 
higher-order correction ${\cal L}_c$ only for $\varphi~\lsim~1$.  The 
curvature perturbation does not exhibit significant variation during the 
contracting phase.  However small-scale modes are enhanced 
around the graceful exit.
\end{figcaption}
\end{center}
\end{figure}

We show in Fig.~\ref{spectrum} the spectra of curvature perturbations
for $p=10^{-3}$ in the case where the correction ${\cal L}_c$ is included
only for $\varphi~\lsim~1$.  We find that the numerical value of the 
spectral tilt is $n_{\cal R}-1 \sim 2$ for $k~\lsim~10^{-4}$, 
which coincides with the analytic estimation (\ref{tilt_po}).
However this estimate is no longer valid for small-scale modes due to the 
negative frequency shift and the instability around the graceful exit.  The 
spectra are highly blue-tilted for $k~\gsim~10^{-4}$ as found in 
Fig.~\ref{spectrum}.  This growth of small-scale fluctuations obviously 
works as the gravitational back-reaction to the background evolution.  
Although we did not consider the effect of the back-reaction here, it is 
certainly of interest to investigate how the background evolution is 
modified around the graceful exit. We have performed the simulations
with various choices of time steps to make sure that the effects we
find are not numerical artifacts. The spectra obtained are independent
of the specific value of the time step.

\begin{figure}
\begin{center}
\singlefig{10cm}{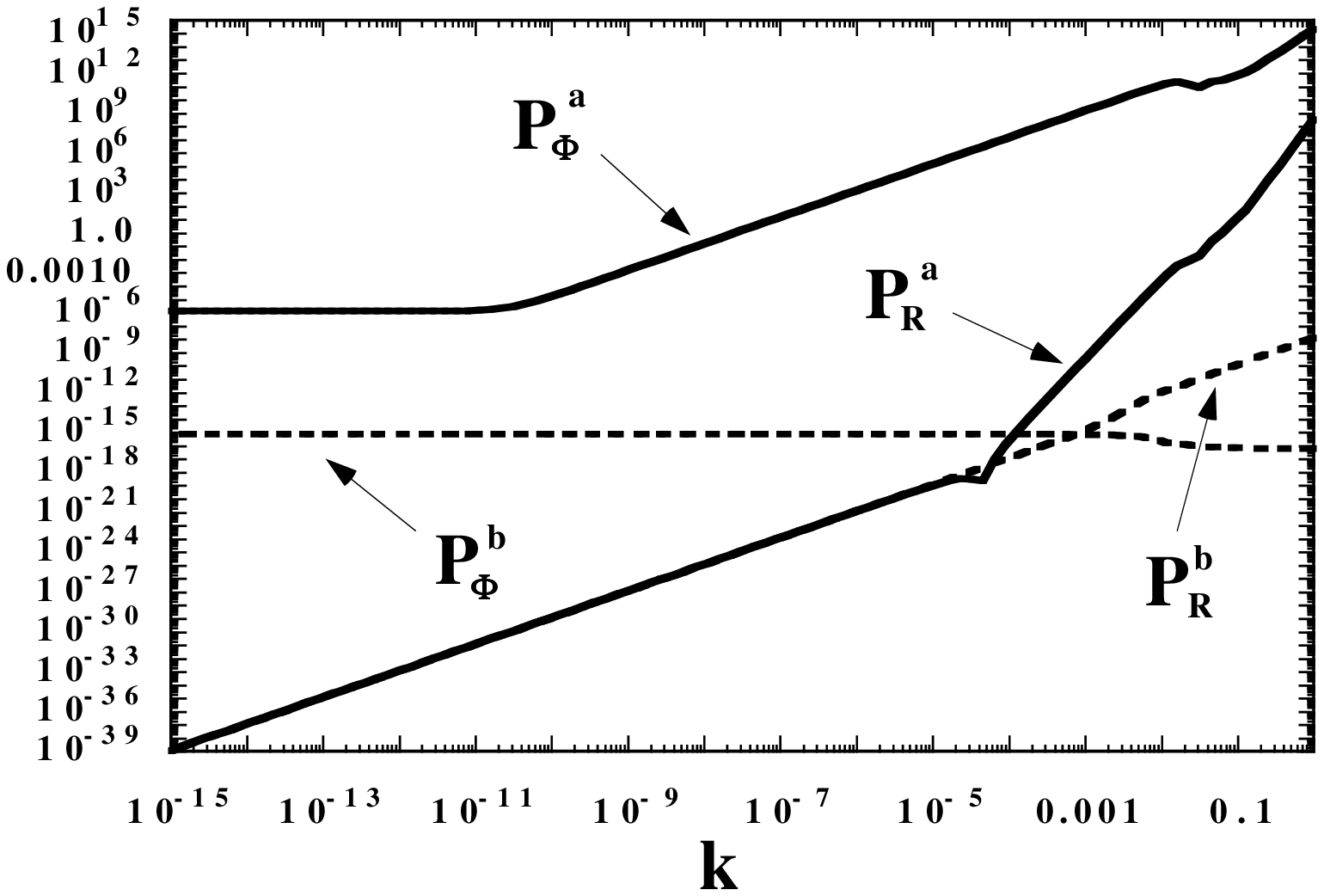}
\begin{figcaption}{spectrum}{10cm}
The final spectra of the curvature perturbation, $P_{\cal R}$, and of
the gravitational potential, $P_{\Phi}$,  in a simulation with $c=-1$, $d=1$, 
$p=10^{-3}$, $C_1=1.0$ and $C_2= -1.0 \times 10^{-3}$ 
with same initial conditions as in Fig.~\ref{speevolution}. 
The superscript ``b'' denotes the quantities {\it before} the bounce 
(at $t_S=300$), while ``a'' indicates the quantities {\it after} 
the bounce (at $t_S=1000$).  
We have included the quantum correction ${\cal L}_c$ for 
values $\varphi~\lsim~1$.  The spectral tilt is $n_{\cal{R}}-1 \sim 2$ for 
$k~\lsim~10^{-4}$, which agrees with the analytic estimation of 
eq.~(\ref{tilt_po}).  For the modes $k > 10^{-4}$, the spectra are highly 
blue-tilted, due to an instability of small-scale modes during the graceful 
exit.  The fluctuations in $\Phi$ are nearly scale-invariant on large scales 
before the bounce.
\end{figcaption}
\end{center}
\end{figure}

In Fig.~\ref{spectrum} we have also plotted the induced fluctuations
of $\Phi_k^E$ in the Einstein frame, determined from the results for ${\cal 
R}_k^S$ in the string frame, using the relation
\begin{equation}
\Phi_k^E \propto \dot{\cal R}_k^S/k^2 \,. 
\label{relphi}
\end{equation}
This corresponds to eq.~(\ref{modi}) in Appendix B, which follows from the 
relation (\ref{Phi}) in the Einstein frame.  Note that this relation is 
valid in the absence of higher curvature corrections to the Lagrangian, and 
will therefore be good at times long before and long after the bounce.  For 
the negative Ekpyrotic potential ($0<p<1/3$), one has $0<\gamma<1/2$ from 
eq.~(\ref{gamma_po}), in which case the second term in eq.~(\ref{dotPsi}) 
completely vanishes.  In this case we have the relation (\ref{Phi3}), 
namely 
\begin{eqnarray} 
\Phi_k^E \propto H_{\nu-1}^{(1, 2)}/k \,.
\label{Phi2}
\end{eqnarray}
Note that $H_{\nu-1}^{(1, 2)}$ can be written by the 
sum of two terms which are proportional to $(k|\eta_S|)^{\nu-1}$ and 
$(k|\eta_S|)^{-\nu+1}$.  

Since $0<\nu<1/2$ for $0<\gamma<1/2$ (i.e., $0<p<1/3$), the term 
proportional to $(k|\eta_S|)^{\nu-1}$ is the growing mode during the 
contracting phase on large scales.  Therefore the spectrum of $\Phi_k^E$ 
before the bounce can be estimated as 
\begin{eqnarray} 
P_{\Phi}^{{\rm b}} \propto k^{2\nu -1} \propto k^{-2\gamma} \propto 
k^{n_{\Phi}-1} \,,
\label{PPhib}
\end{eqnarray}
from which we have
\begin{eqnarray} 
n_{\Phi}-1=-\frac{2p}{1-p} \,.
\label{Phispe}
\end{eqnarray}
Then we have a scale-invariant spectrum before the bounce
for $p \sim 0$, as first pointed out in \cite{Khoury:2001zk}.  
This agrees with our numerical result shown in 
Fig.~\ref{spectrum}.

The term proportional to 
$(k|\eta_S|)^{\nu-1}$, however, decays after the graceful exit as long as 
$\nu<1$.
The dominant mode in $\Phi_k^E$ long after the bounce 
is described by the term $(k|\eta|)^{-\nu+1}$, in which case the spectrum 
of $\Phi$ is written as 
\begin{eqnarray} 
P_{\Phi}^{{\rm a}} \propto k^{3-2\nu} \,.
\label{PPhia}
\end{eqnarray}
{}From Fig.~\ref{spectrum} we find  that the spectrum of $\Phi$ 
is blue-tilted with $n_{\Phi} \sim 3$ for $k~\gsim~10^{-10}$ 
(small-scale modes for $k~\gsim~10^{-2}$ exhibit larger blue tilt with 
$n_{\Phi}>3$).  This corresponds to the value $\nu \sim 1/2$ in 
eq.~(\ref{PPhia}) after the bounce.

Our numerical calculations show that large-scale modes 
with $k~\lsim~10^{-10}$ do not exhibit such a blue spectrum.  
This can be understood that the term proportional to
$(k|\eta_S|)^{\nu-1}$ which is dominant in the contracting phase
does not become smaller than the one proportional to
$(k|\eta_S|)^{-\nu+1}$ in the expanding branch for very small $k$, unless we 
evolve the fluctuations until long after the bounce.  However it is 
rather difficult to follow such large amount of time numerically.  In 
addition the second term in eq.~(\ref{dotPsi}) is not numerically 
negligible relative to the first term for these large-scale modes due to 
the modification of the equation of state after the bounce [when 
$\gamma>1/2$ the second term in eq.~(\ref{dotPsi}) is nonvanishing as found 
by eq.~(\ref{re})].  Nevertheless, we expect that the term proportional to 
$(k|\eta_S|)^{-\nu+1}$ in the first term in eq.~(\ref{dotPsi}) eventually 
dominates long after the bounce, in which case the spectrum is given by 
eq.~(\ref{PPhia}).  Therefore the final spectrum of $\Phi$ is not generally 
scale-invariant.  The spectral index is dependent on the evolution of the 
background after the bounce (i.e., $\gamma$).  In this sense including 
radiation is necessary in order to evaluate the spectrum of $\Phi$ in 
realistic cases where the solution connects to our Friedmann branch.

{}From Fig.~\ref{spectrdecay} we find that 
the amplitude of $\Phi$ decreases after the bounce, thus showing
that the dominant pre-bounce mode of $\Phi$ couples exclusively
to the decaying mode of $\Phi$ after the bounce, as derived in
\cite{Brandenberger:2001bs} using matching conditions on a
constant scalar field hypersurface.

\begin{figure}
\begin{center}
\singlefig{10cm}{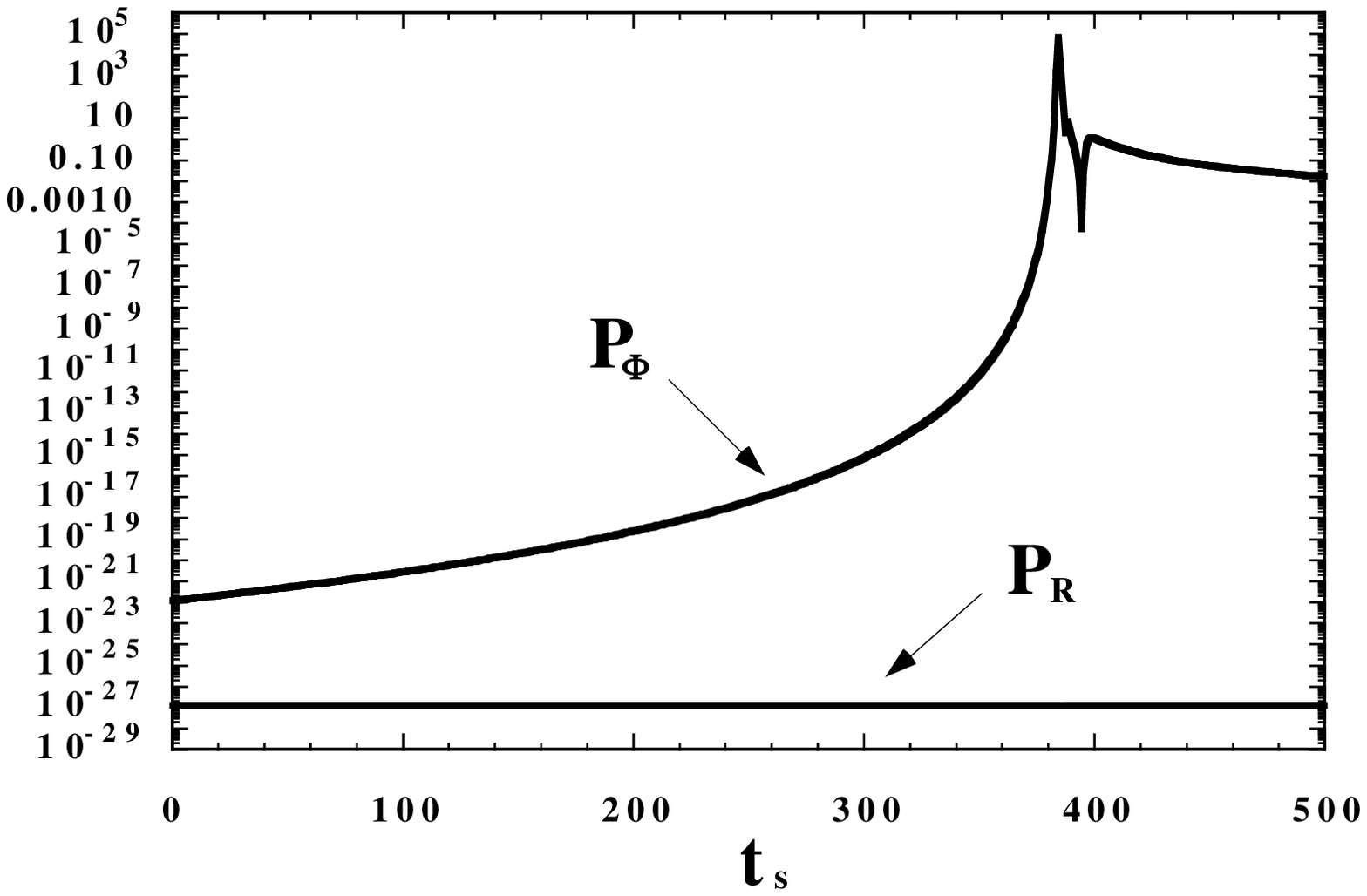}
\begin{figcaption}{spectrdecay}{10cm}
The evolution of the curvature perturbation, $P_{\cal R}$, and of
the gravitational potential, $P_{\Phi}$ for the fluctuation mode
corresponding to $k = 10^{-9}$. 
The model parameters and initial conditions are the same as
in Fig.~\ref{spectrum}.
The amplitude of the gravitational potential near the bounce when 
the higher derivative terms dominate cannot be trusted since 
$\Phi_k^E$ is computed from ${\cal R}_k^S$ via 
the equation (\ref{relphi}) which is only valid 
in the absence of such higher derivative terms. 
As follows from this plot, the dominant growing mode of $\Phi_k^E$ 
during the period of contraction couples
only to the post-bounce decaying mode. At the time of the bounce,
the curve for $\Phi$ is dominated by numerical noise. However, since
$\Phi$ is computed at each time separately from the value of ${\cal R}$
at that time, this does not introduce numerical errors in the late
time values of $\Phi$.
\end{figcaption}
\end{center}
\end{figure}

Eqs.~(\ref{PR}) and  (\ref{gamma_po}) indicate that a scale-invariant 
spectrum may be 
obtained for $p=2/3$ for the modes which are enhanced during the bouncing 
phase \cite{Finelli:2001sr}.  In order to obtain such a spectrum, the 
exponential potential (positive in this case) is required to dominate the 
higher-order term except around the graceful exit.  However we have found 
that, for some likely initial conditions, the field $\varphi$ bounces back 
toward larger $\varphi$ before the higher-order correction begins to work.  
In language appropriate to Ekpyrotic cosmology, this means that the branes 
never collide.  If the higher-order term always dominates compared to the 
positive exponential potential, we have blue-tilted spectra (\ref{tilt1}).

 \section{Inclusion of higher-order corrections in the Einstein
 frame \\
 --modulus driven case} 

In this section we consider adding higher derivative terms
defined in the Einstein frame. We add a Gauss-Bonnet term
proportional to $R_{\rm GB}^2$ multiplied by a function of
the modulus field $\varphi$ to the action. Such a term arises
as the one loop correction in the context of orbifold compactifications
of the heterotic superstring \cite{Antoniadis:1992rq}. Since the
initial version of Ekpyrotic cosmology \cite{Khoury:2001wf} is
based on an orbifold compactification of a theory dual to heterotic
superstring theory, the correction terms used in this section are
well motivated in the context of the scenario of \cite{Khoury:2001wf}.
Indeed, it was found in Ref.~\cite{Antoniadis:1993jc} (see also
\cite{Rizos:1993rt}-\cite{Alekseev:eh})
that the inclusion of the Gauss-Bonnet 
term coupled to a modulus field in the Einstein frame leads to 
the possibility of obtaining nonsingular solutions. In the work of
\cite{Antoniadis:1993jc}, the potential for the modulus field was
taken to vanish. In this section we will include an 
exponential potential.\footnote{The authors of 
ref.~\cite{Alekseev:eh} analyzed nonsingular cosmological
solutions in the presence of some positive potentials (not the  
exponential potential).}
More specifically, the correction Lagrangian we consider here
corresponds to [in the notation of (\ref{lag}) and (\ref{lagalpha})]  
$f=R$, $\omega=1$, $c=-1$, $d=0$,
$\xi(\varphi)={\rm ln}[2e^{\varphi}\eta^4(ie^{\varphi})]$ with 
$\eta(ie^{\varphi})$ being the Dedekind $\eta$-function 
\cite{Antoniadis:1993jc}.  
Here $\xi(\varphi)$ is approximately given by 
\begin{eqnarray}
\xi(\varphi) \simeq -\frac{\pi}{3}
\left(e^{\varphi}+e^{-\varphi}\right)\,.
\label{xiapprox}
\end{eqnarray}
The sign of $\lambda$ is chosen to be positive, which is 
different from the one discussed in the previous 
section. Note also that even though
the coefficient $\xi(\varphi)$ becomes large at large brane
separation (large negative values of the dilaton in the case
of PBB cosmology), this increase is outweighed by the falloff
of the curvature invariant, as in the case of the model considered
in the previous section. Thus, in Ekpyrotic cosmology
the correction terms are expected  to
become important only in the high curvature region.  

Let us analyze the one-field system of a modulus $\varphi$, 
keeping the dilaton fixed. We will consider solutions starting
in an asymptotically flat region and beginning in the expanding
branch. We have not found solutions which begin in a contracting
phase and undergo a successful bounce. However, note that in
the original Ekpyrotic scenario of \cite{Khoury:2001wf}, the scale
factor on the orbifold fixed plane corresponding to our 
four-dimensional space-time corresponds to an initially asymptotically 
flat region, and is always expanding. Thus, the solutions found here might be applicable
to a version of Ekpyrotic cosmology formulated entirely in terms of
physics on the orbifold fixed plane. 

 \subsection{Background evolution} 

As was discussed in \cite{Antoniadis:1993jc}, when $V_E=0$ the PBB 
singularity can be avoided for positive values of $\lambda$ when the
$\alpha'$ corrections introduced above are taken into account.  
The sign of $\lambda$ is crucial 
for the existence of nonsingular cosmological solutions.
For negative values of $\lambda$, the $\alpha'$ corrections do not help to 
lead to a successful graceful exit, as was analyzed in 
ref.~\cite{Toporensky:2002ta}. 

In the absence of the Ekpyrotic potential, 
the background evolution for $t_E<0$ is given by \cite{Kawai:1998bn} 
\begin{eqnarray}
a_E \simeq a_0,~~~H_E=\frac{H_0}{t_E^2},~~~
\dot{\varphi}=\frac{5}{t_E},
\label{backevo}
\end{eqnarray}
where $a_0$ and $H_0$ ($>0$) are constants.
The Gauss-Bonnet term leads to a  
violation of the null energy condition ($\rho_E+p_E<0$)
at sufficiently large curvatures and thus enables a graceful exit.
If we start in an expanding branch (contrary to the
spirit of PBB and Ekpyrotic cosmology), this leads to 
a super-inflationary solution ($\dot{H}_E>0$) 
until a ``graceful exit'' (see Fig.~\ref{modulus}).
The Universe is initially 
expanding very slowly  with a nearly constant scale factor.
After the Hubble parameter reaches its peak value 
$H_E=H_{\rm max}$, the system connects to a Friedmann-like Universe with 
$H_E \simeq 1/(3t_E)$, $a_E \propto t_E^{1/3}$, and 
$\varphi \propto -{\rm ln} t_E$.
 
If the Ekpyrotic potential is present, the situation is 
quite different.  
We have adopted the potential (\ref{einpoten}) 
for $\varphi>0$ and $V_E=0$ for $\varphi<0$. Once again,
we start in an expanding phase. Initially, the potential term
is not important and the Universe evolves in a super-inflationary
trajectory until a graceful exit after which the Hubble expansion
rate begins to decrease. When $p<1/3$, corresponding to the case of a 
negative exponential potential, then as $\varphi \to 0$, the potential becomes
important and leads to a change in sign of $H_E$.
We find that the system enters a stage of slow contraction 
[see the case (ii) of Fig.~\ref{modulus}]. 
Note that in Fig.~\ref{modulus} $\dot{H}_E$ changes the sign twice.
After the negative Hubble peak, the Hubble rate begins to grow toward 
$H_E \to -0$ 
without changing sign. Then the system enters a very slow contracting phase
with a nearly constant scale factor. In this stage the field $\varphi$ 
evolves rapidly toward large negative values. 
In the presence of negative exponential potential ($p<1/3$) 
we have found that 
the contracting stage eventually appears even when $V_0$ is small.
 
\begin{figure}
\begin{center}
\singlefig{15cm}{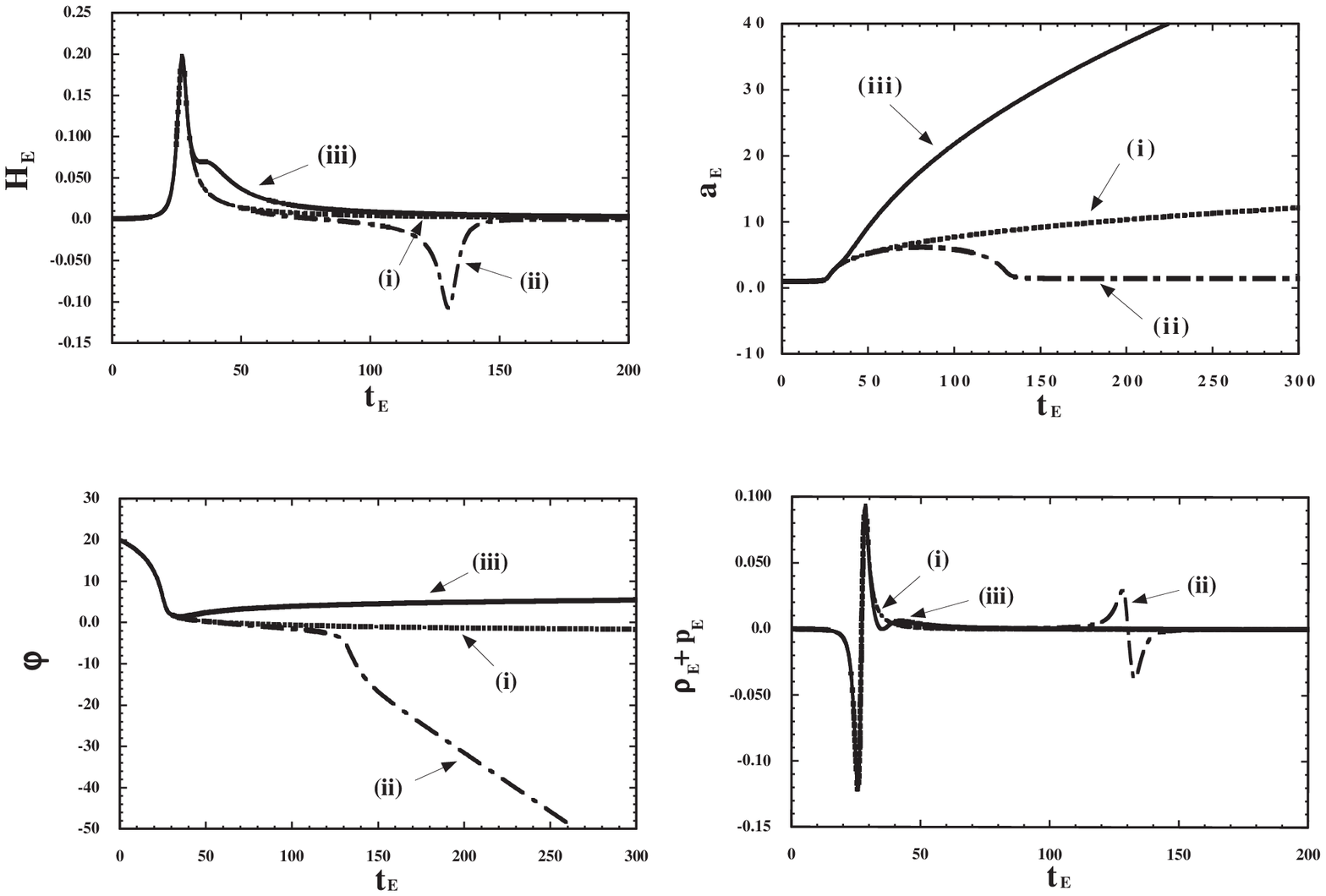}
\begin{figcaption}{modulus}{15cm}
The evolution of $H_E$, $a_E$, $\varphi$, and 
$\rho_E+p_E$ in the modulus-driven case 
with $c=-1$, $d=0$.
We choose initial conditions $\varphi=20$ and
$H=5.087 \times 10^{-4}$.
Each case corresponds to (i) $V_0=0$ with $p=0.1$,
(ii) $V_0=0.01p(1-3p)$ with $p=0.1$,
(iii) $V_0=p(1-3p)$ with $p=0.5$.
\end{figcaption}
\end{center}
\end{figure}

When $p>1/3$ there exists a positive potential barrier as the field $\varphi$
approaches zero. The case (iii) in Fig.~\ref{modulus} corresponds to 
$p=1/2$ with $V_0=p(1-3p)$. The effect of the positive potential
is important around $\varphi \sim 0$, which works to return the field back 
toward larger $\varphi$. After the graceful exit the Hubble rate is always positive
with slowly changing $\varphi$. The scale factor evolves as a power-law    
($a \propto t^p$) due to a positive exponential potential.
The $p>1/3$ case provides us with reasonable nonsingular 
cosmological solutions.
Nevertheless we need to caution that  these nonsingular solutions
are different from the bouncing ones where the contraction of 
the universe occurs before the graceful exit.

One may argue that the bouncing trajectories may be found 
by including the correction ${\cal L}_c$ only around $\varphi \sim 0$.
However we have numerically found that this is not the case.
The super-inflationary evolution characterized by eq.~(\ref{backevo}) is   
typically required for the construction of nonsingular solutions 
in the present scenario. 

 \subsection{Density perturbations}

When the Gauss-Bonnet term is dominant relative to the Ekpyrotic potential, 
the spectra of density perturbations can be analyzed as in the case of
the zero potential ($p=0$ or $p=1/3$).
In this case the evolution of the background during the phase of 
modulus-driven 
inflation is given by eq.~(\ref{backevo}), thereby leading to 
$\dot{\xi}(\varphi) \simeq
-(\pi/3)\dot{\varphi} e^{\varphi} \propto (-t_E)^4$. 
Making use of this relation together with 
eq.~(\ref{Qs2}), we find the evolution of $Q$ and $z$ as 
\begin{eqnarray}
Q \propto (-t_E)^2,~~~~z \propto (-t_E) \propto (-\eta_E)\,.
\label{Qz}
\end{eqnarray}
This means that $\gamma=1$ in eq.~(\ref{z}), in which case
curvature perturbations are enhanced on super-Hubble scales during 
super-inflation (${\cal R}_k \propto (-\eta_E)^{-1}$)
due to the growth of the second term in eq.~(\ref{Rk})
\cite{Kawai:1999pw}.  
We show in Fig.~\ref{modu_evo} the evolution of 
curvature perturbations in the case of zero potential ($p=1/3$) for two 
different modes ($k=10^{-3}$ and $k=10^{-1}$). 
We find that curvature perturbations are amplified before the graceful exit.

In order to obtain the spectral tilt of density perturbations, we have to 
caution that the function $s$ defined by eq.~(\ref{Qs}) is a time-varying 
function and is proportional to $(-t_E)$.
Therefore the formula (\ref{ind}) can not be directly applied. 
Instead one is required to consider the evolution 
equation for curvature perturbations: 
\begin{eqnarray}
\ddot{{\cal R}}_k+\frac{2}{t_E}\dot{{\cal R}}_k-\alpha
\frac{k^2}{a_0^2} t_E {\cal R}_k=0\,,
\label{dR_k}
\end{eqnarray}
where $\alpha~(>0)$ is a constant that depends on $H_0$ 
in eq.~(\ref{backevo}).  
The solution of this equation is written in terms of the Bessel functions 
\begin{eqnarray}
{\cal R}_k=(-t_E)^{-1/2} \left[c_1 J_{-1/3}(x)+c_2J_{1/3}(x)\right]\,,
\label{Bessel}
\end{eqnarray}
where $x \equiv \frac23 \sqrt{\alpha} \frac{k}{a_0}(-t_E)^{3/2}$.  
Notice that this solution asymptotically approaches the 
Minkowski vacuum for $x \to \infty$.  
Since $J_{\pm 1/3} (x) \propto k^{\pm 
1/3}$ in the $x \to 0$ limit, the spectrum of large-scale curvature 
perturbation is proportional to  
$P_{{\cal R}} \propto k^{7/3}$.
Therefore the spectral index is  
\begin{eqnarray}
n_{\cal R}-1=\frac73\,,
\label{PowerM}
\end{eqnarray}
which is a blue-tilted spectrum.

In the absence of the Ekpyrotic potential, the evolution of the background 
in the asymptotic future is given by $\varphi \sim \sqrt{3/2}\,{\rm ln} t_E$,
$H_E \propto 1/(3t_E)$, and $a \propto t_E^{1/3}$.  Therefore one has 
$z \propto t_E^{1/3} \propto \eta_E^{1/2}$ in eq.~(\ref{Rk}), 
in which case curvature perturbations exhibit logarithmic growth,
\begin{eqnarray}
{\cal R} \propto {\rm ln}~\eta_E \,.
\label{log}
\end{eqnarray}
This indicates that the second term in eq.~(\ref{Rk}), which we call 
``D-mode'', dominates even after the graceful 
exit.  In the case where the D- mode decays after the graceful exit, 
the surviving spectra observed in an expanding Universe should correspond 
to the first term in eq.~(\ref{Rk}) [``C-mode''].  In the present model, 
however, the D-mode survives in an expanding branch.  Therefore the 
spectrum of the curvature perturbation during super-inflation can be 
preserved even after the graceful exit.  In fact, the numerical value of the 
final spectral tilts of ${\cal R}$ are found to be $n_{\cal R}-1 \sim 2.3$ 
for the modes $k \ll 1$ (see Fig.~\ref{modu_spectra}). 
This agrees well with the analytic result (\ref{PowerM}).

\begin{figure}
\begin{center}
\singlefig{10cm}{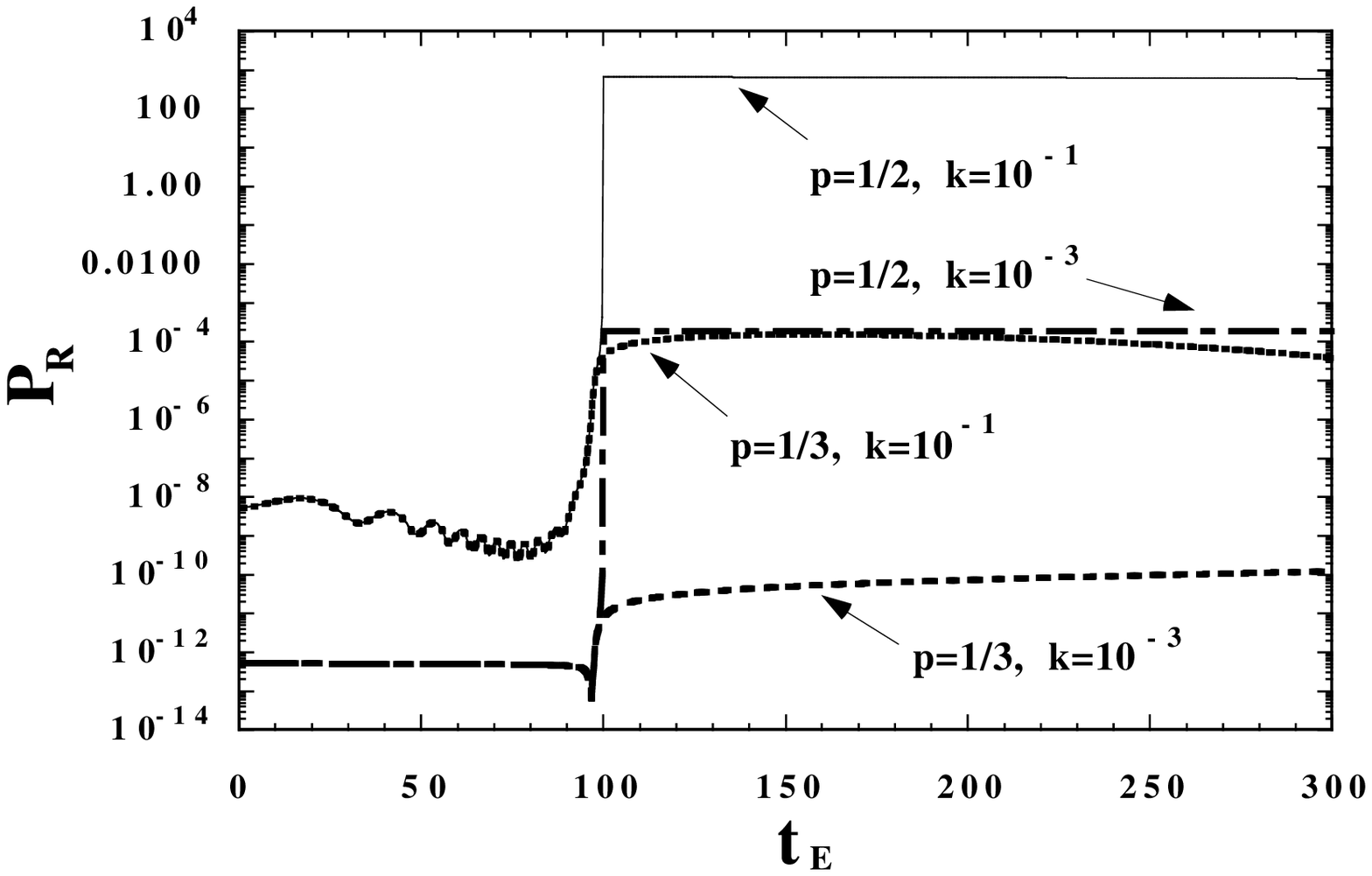}
\begin{figcaption}{modu_evo}{10cm}
The evolution of the spectra of curvature perturbations in the modulus-driven 
case ($c=-1$ and $d=0$) for $(p, k)=(1/3, 10^{-1}), (1/3, 10^{-3}), (1/2, 10^{-1}),
(1/2, 10^{-3})$.  
We choose initial conditions $\varphi=23.888$ and
$H=4.158 \times 10^{-4}$.
Note that the $p=1/2$ case corresponds to the positive
exponential potential while the zero-potential corresponds to $p=1/3$.  Around 
the graceful exit curvature perturbations exhibit rapid growth especially 
when the potential is positive $(p>1/3)$.
\end{figcaption}
\end{center}
\end{figure}

\begin{figure}
\begin{center}
\singlefig{10cm}{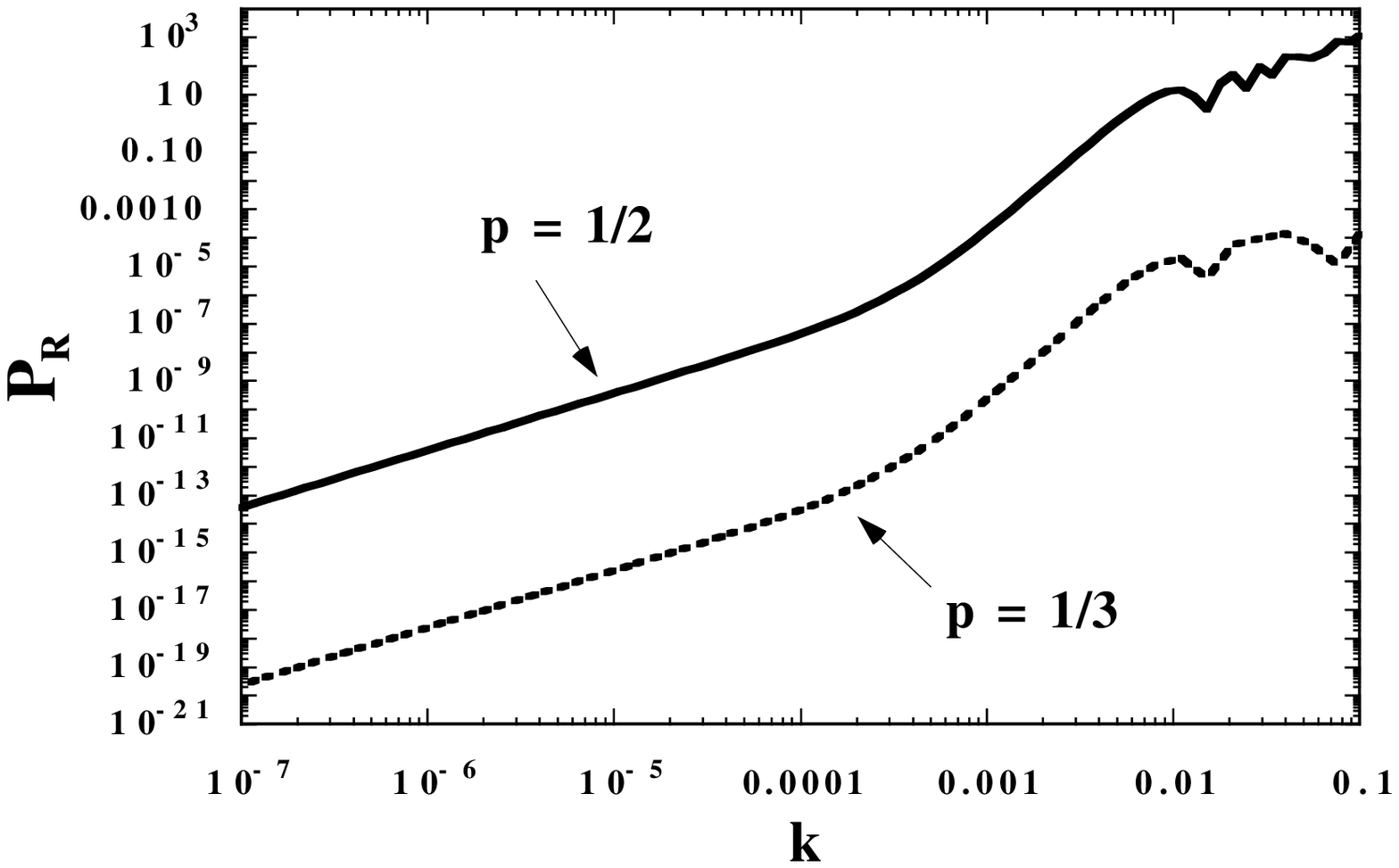}
\begin{figcaption}{modu_spectra}{10cm}
The final spectra of curvature perturbations
in the modulus-driven case ($c=-1$ and $d=0$) for $p=1/3$ and $p=1/2$.
The initial conditions are the same as in Fig.~\ref{modu_evo}.
When the positive potential is present ($p>1/3$), the amplitude of the spectrum is
larger than in the case of the zero-potential. 
We numerically find that the spectral tilt is 
$n_{\cal R}-1 \sim 2.3$ for $k \ll 1$, 
which agrees well with the analytic estimation, $n_{\cal R}-1=7/3$.
\end{figcaption}
\end{center}
\end{figure}

When a positive Ekpyrotic potential is present ($p>1/3$), the dynamics of the 
system is more unstable around the graceful exit.  This leads to the 
violent growth of curvature perturbations when the field bounces back due to 
the potential barrier.  This threatens the viability of the cosmological 
perturbation theory around the graceful exit.  Nevertheless the perturbations
are not singular as long as the background is smoothly joined to the 
expanding branch.  We have numerically evaluated the power spectra of 
${\cal R}$ for the modes which left the horizon during super-inflation. 
Although the amplitude is larger compared to the case of the zero-potential 
($p=1/3$), the final spectral tilts are similar for large-scale modes
($k \ll 1$), see Fig.~\ref{modu_spectra}.  Again the final spectra are 
found to be blue-tilted.  We should also mention that the frequency 
shift $s$ becomes negative for the Hubble rate which is larger than unity 
around the graceful exit \cite{Kawai:1999pw}. 
In this case the small-scale modes show
exponential instability as we pointed out in the dilaton-driven case.
The negative Ekpyrotic potential ($p<1/3$) is not worth 
studying, since this case does not connect to the expanding branch as 
analyzed in the previous subsection.

Finally, we should mention that we have neglected the effect of radiation 
in all our analysis.  
However this is expected to appear at some moment of time.
This can also alter the final spectra of curvature perturbations due to the 
dominance of the C-mode in eq.~(\ref{Rk}).
We leave to future work about these realistic situations.

 \section{Discussions and Open Issues}

We have studied the effects of higher derivative terms in the
joint gravitational and matter action for theories motivated
by Pre-Big-Bang and Ekpyrotic cosmology with a single scalar
matter field with an exponential potential. Applied to
PBB cosmology, our model corresponds to a theory with an
exponential potential for the dilaton. In the language of
the initial version of Ekpyrotic cosmology \cite{Khoury:2001wf}, 
our scalar field is the modulus field
corresponding to the separation of the bulk brane from our
orbifold fixed plane, in the second version \cite{Khoury:2001bz}
and in its cyclic version \cite{Steinhardt:2001vw,Steinhardt:2001st}
the field is the radius of the extra spatial dimension. The
higher derivative terms introduced are the leading string and
quantum corrections to the low energy effective action of
string theory.

When applying the correction terms in the string frame, and
for suitable choices of the coefficients of the higher-order 
corrections we find
nonsingular cosmological solutions which in the
Einstein frame correspond to bouncing Universes. We thus find that
higher derivative terms can smooth out the singularities in PBB
and Ekpyrotic cosmology and lead to a graceful exit (in the
language of PBB cosmology) or a nonsingular bounce (in the
language of Ekpyrotic cosmology). We have thus generalized the
results of \cite{Gasperini:1996fu} to models with exponential
scalar field potentials.

We have studied the evolution of fluctuations in our nonsingular
bouncing cosmologies. This analysis is not plagued by the
matching ambiguities inherent to analyses where the
contracting and expanding cosmologies are matched across
a singular space-time surface. For all potentials with $0 < p \ll 1$ we 
find a blue spectrum of curvature fluctuations. The precise
spectral index depends, as expected, on whether the higher
derivative correction terms are important at times when the
scales on which we compute the fluctuation spectrum exit the Hubble
radius during the phase of contraction. 

If the higher derivative terms are not dominant when the scales
exit the Hubble radius, the index of the spectrum agrees with what
is obtained by applying the general relativistic matching conditions
on a uniform density hypersurface 
\cite{Brandenberger:2001bs,Hwang:2001ga,Tsujikawa:2001ad}. 
The only difference is an instability of small-scale fluctuation modes during
the bounce (see also \cite{Kawai:1999pw}) which leads to a further
steepening of the spectrum. Our result implies that the growing
mode of $\Phi$ during the contracting phase which is 
scale-invariant for $0 < p \ll 1$ is effectively uncoupled with the
dominant constant mode of $\Phi$ in the expanding phase, a result
obtained in the context of matching conditions in \cite{BGGMV} (in
the case of PBB cosmology) and in 
\cite{Brandenberger:2001bs,Hwang:2001ga} for the Ekpyrotic scenario.
If the higher derivative terms dominate when scales of interest
exit the Hubble radius, then the spectrum is blue with a slope
of $n_{\cal R}-1 \simeq 2.26$. Note that our result implies that it is
the curvature fluctuation ${\cal R}$ (more precisely the variable
${\tilde \zeta}$ originally introduced by Bardeen in \cite{Bardeen:1988hy}
and used in \cite{Brandenberger:2001bs}, which equals ${\cal R}$ up to
terms which are suppressed by $k^2$ for large-scale fluctuations)
which is effectively conserved for large-scale perturbations across
the bounce.
 
We have also studied nonsingular cosmological models obtained 
by adding a Gauss-Bonnet term (defined in the Einstein
frame) multiplied by a suitably chosen
function of the single scalar matter field in the problem (a
modulus field). Once again, we have included an exponential
potential for the modulus field. Although we do not find bouncing
cosmologies, we find interesting nonsingular cosmological solutions
which begin in an asymptotically flat region, undergo super-exponential
inflation followed by a graceful exit to a phase with decreasing Hubble
radius. In the presence of a negative exponential potential
($0 < p < 1/3$), the solutions reach a maximal radius and 
begin to contract as the field crosses $\varphi=0$.
 During this period of
contraction, the Hubble
parameter remains finite. Such solutions might be applicable to
Ekpyrotic Universe models formulated in terms of physics on the
four-dimensional orbifold fixed plane corresponding to our visible
space-time.  
When the potential is positive 
($p>1/3$), the modulus $\varphi$ bounces back around the brane collision 
toward larger $\varphi$ due to the barrier of the positive potential.  
Although singularities can be avoided around $\varphi \sim 0$, this 
model does 
not correspond to bouncing solutions where the contraction of the Universe
occurs due to the Ekpyrotic potential before the graceful exit.

We have also studied the spectrum of curvature fluctuations ${\cal R}$
in this modulus-driven cosmology. 
When higher-order corrections are important before the bounce
(as it must be for the existence of nonsingular solutions),
one has $n_{\cal R}-1 \sim 7/3$. 
This result is again
in agreement with what can be obtained by neglecting the graceful
exit and matching two Einstein Universes at a constant density
hypersurface. 

Note that we have chosen to evolve the curvature perturbation 
${\cal R}$ {on comoving hypersurfaces}, and found that the spectral 
index is given by 
(\ref{tilt_po}) when the Ekpyrotic potential is dominant.  Note that this
spectrum in the $0 < p </ 1/3$ case comes from the C-mode in eq.~(\ref{Rk}), 
which is blue-tilted for very slow contraction ($0 < p \ll 1$).  Since the 
large-scale D-modes are enhanced for $1/3<p<1$ during the 
contracting phase, 
the spectrum of ${\cal R}$ will be scale-invariant for $p \sim 2/3$ {\it 
right after} the bounce \cite{Finelli:2001sr}.  

If we follow instead the gravitational potential $\Phi$ 
in the longitudinal gauge, its spectral index generated during the 
collapsing phase is estimated as eq.~(\ref{Phispe}), which is 
different from that of ${\cal R}$ [see eq.~(\ref{tilt_po})].
When $p \sim 0$, corresponding to a very 
slow contraction, the growing mode (D-mode) of $\Phi$ is approximately
scale-invariant.
The authors of ref.~\cite{Khoury:2001zk,Durrer:2002jn} 
claimed that a scale-invariant spectrum of the dominant post-bounce
mode of $\Phi$ would inherit this scale-invariant spectrum.

However, we know that when 
computed at late times long after the bounce, in an expanding universe, the 
spectra of ${\cal R}$ and $\Phi$ must be identical. Thus, given
our results concerning the spectrum of ${\cal R}$, we know that
the spectrum of $\Phi$ long after the bounce cannot be scale-invariant.
Our numerical simulations show that the contribution from the
pre-bounce D-mode decays 
after the system enters the expanding branch, and thus shows that
the pre-bounce growing mode of $\Phi$ couples exclusively to the
post-bounce decaying mode. This results in 
a blue-tilted spectrum of $\Phi$ when evaluated long after the bounce 
(see Fig.~\ref{spectrum}). 
For very large-scales with $k~\lsim~10^{-10}$, we need to solve
the equation of fluctuations up to sufficient amount of time
in order to find the complete decay of the D-mode relative to the 
C-mode. In addition the second term in eq.~(\ref{dotPsi}) is not numerically 
negligible for very small $k$ when $\gamma$ is greater than $1/2$.  
Nevertheless the term proportional to $(k|\eta_S|)^{-\nu+1}$ in the first 
term in eq.~(\ref{dotPsi}) eventually dominates long after the bounce 
($\eta_S \to \infty$), thereby yielding the spectrum (\ref{PPhia}).  
Therefore the spectrum of $\Phi$ long after the bounce is not generally 
expected to be scale-invariant, whose spectral index depends on the 
evolution of the background in an expanding branch.

Since near the bounce the magnitudes of the two modes of $\Phi$ and
${\cal R}$ differ by such a large ratio, we must worry about the
possibility of numerical errors. In particular, if one were to follow
the evolution equation for $\Phi$, it would be difficult to ensure that
numerical noise does not lead to an artificial coupling between the
pre-bounce growing mode and the post-bounce dominant (constant) mode.
We have checked that our results do not seem to suffer from a similar
problem by repeating the simulations with different values of the 
time step, $\Delta t$.
We did not find any dependence of the results within the range of time steps
we have chosen ($10^{-5} \le \Delta t \le 10^{-3}$).

Let us compare our findings to results which have already appeared
in the literature. As mentioned repeatedly, our results concerning
the spectrum of fluctuations obtained in the classes of nonsingular bouncing
Universe models considered in this paper agree with the results of
\cite{Brandenberger:2001bs,Hwang:2001ga,Tsujikawa:2001ad} 
obtained when removing the higher derivative correction
terms (thus going back to a singular background) and matching the
fluctuations on a constant scalar field matching surface. The
results imply that the growing mode of $\Phi$ in the contracting
phase does not source the post-bounce dominant mode of $\Phi$.
Our results thus indicate that the conjecture of 
\cite{Khoury:2001zk,Durrer:2002jn}, namely
that the growing mode of $\Phi$ in the contracting phase (which in
Ekpyrotic cosmology has a scale-invariant spectrum) should {\it generically}
determine the amplitude and spectrum of the dominant mode of $\Phi$
in the post-bounce phase, is not valid.  As emphasized in \cite{Martin:2001ue} 
and \cite{Durrer:2002jn}, in the case of a singular background the spectrum 
of fluctuations in the expanding phase depends sensitively on the details 
of the matching conditions used.  Since we have only used one class of ways 
to smooth out the singularity, the sensitive dependence on the matching 
surface might not have been completely eliminated, but might find itself 
reflected in a sensitive dependence of the final spectrum on the specific 
form of the correction terms in the action.  We leave the study of this 
issue to future work.

Our work indicates that it is difficult to obtain a scale-invariant spectrum 
of curvature fluctuations for a single field PBB or Ekpyrotic cosmology.  
However, in the case of Ekpyrotic cosmology there is the intriguing fact 
that the growing mode of the gravitational potential $\Phi$ during the 
phase of contraction has a scale-invariant spectrum.  To obtain a 
scale-invariant spectrum of $\Phi$ and thus also of the curvature 
fluctuation ${\cal R}$ at late times in the expanding phase, a suggestion 
\cite{Khoury:2001zk,Durrer:2002jn} was to non-trivially connect the growing 
mode of $\Phi$ during the contracting phase with the constant mode in the 
expanding phase.  We have shown that this does not occur in the single 
field case with our choice of correction terms to the action (needed to 
obtain a nonsingular bounce).

Note that there are examples where a large growth of $\Phi$ during the
phase of contraction persists after the bounce (see e.g. 
\cite{Finelli:2001sr,Peter:2002cn,Gordon:2002jw}). A criterion for when
this occurs has been proposed recently in \cite{Gordon:2002jw}. 
The condition is that the relative variation of 
${\cal R}$ over a Hubble time scale should be appreciable,  
i.e. the following relation 
\cite{Gordon:2002jw}
\be
\frac{\dot {\cal R}}{H {\cal R}} \gg  1\,,
\label{inequality}
\ee
should hold close to or right at the bounce. We use eq.~(\ref{Rk}) and 
restrict to the case of an exponential potential, 
in which case one has
\begin{eqnarray}
\frac{\dot {\cal R}}{H {\cal R}}=
\frac{D_k (1-p) \eta}{p a^2 (C_k + D_k (-\eta)^{1-2\gamma}) } = 
\cases{ \frac{D_k (1-p)}{p C_k} (-\eta)^{\frac{1-3p}{1-p}} 
& for $0<p<1/3$ \,,
\cr \frac{1-p}{p} & for $1/3<p<1$ \,.
\cr } 
\label{evidence}
\end{eqnarray}
The marginal case $p=1/3$ (important for PBB and also for Ekpyrotic
scenario, in which the potential disappears close to the bounce) should be
treated separately
and leads to:
\be
\frac{\dot {\cal R}}{H {\cal R}} = \frac{2 D_k}{C_k + D_k \ln
(-k\eta)}\,.
\label{evidence2}
\ee 

Surprisingly enough, both the results (\ref{evidence}) and
(\ref{evidence2}) indicate that $\Phi$ could {\em never} match to 
${\cal R}$ nontrivially.
For $p \le 1/3$, $\dot {\cal R}/H {\cal R} \rightarrow 0$ as
$\eta \rightarrow 0$. For $1/3 < p <1$,  $\dot {\cal R}/{H {\cal
R}} \rightarrow (1-p)/p \sim {\cal O}(1)$ as $\eta \rightarrow 0$. This
latter case is interpreted as a variation of ${\cal R}$ rather than a
change induced by $\Phi$. Interestingly enough, if one takes seriously the
ratio (\ref{inequality}), the singularity at the bounce (i.e., if
$\eta=0$ is reached or not) does not matter in the impossibility
of matching $\Phi$ to ${\cal R}$.  

Recently several authors \cite{Peter:2002cn,Gordon:2002jw} considered 
models of a bouncing Universe (realized in \cite{Peter:2002cn} by introducing 
matter violating the weak energy condition and in \cite{Gordon:2002jw} by 
making use of spatial curvature in the background metric) in which ${\cal 
R}$ grows dramatically across the bounce and there is a coupling between 
the growing mode of $\Phi$ in the contracting phase and the dominant mode 
of $\Phi$ in the expanding phase.  In this case, it may be possible to 
obtain a scale-invariant spectrum, as already realized in 
\cite{Finelli:2001sr}.

Although we have concentrated on the density perturbation in the single field 
scenario, the situation can be changed by taking into account a second scalar 
field \cite{Finelli:2001sr,Notari:2002yc,Finelli:2002we}.  The system of 
multi-component scalar fields generally induces isocurvature perturbations, 
which can be the source of adiabatic perturbations.  In such a case the 
relation (\ref{inequality}) could be satisfied, since isocurvature 
perturbations act as source term for $\dot {\cal R}$ in addition to $\Phi$.  
In fact the authors of ref.~\cite{Notari:2002yc} considered a specific 
two-field system with a brane-modulus $\varphi$ and a dilaton $\chi$.  When 
the dilaton has a negative exponential potential with a suppressed 
Ekpyrotic potential for $\varphi$, the entropy ``field'' perturbation can 
be scale-invariant if the model parameters are fine-tuned 
\cite{Notari:2002yc}.  It was also pointed out in 
ref.~\cite{Finelli:2001sr} that the quantum fluctuation of a light scalar 
field (with a non-canonical kinetic term as studied in 
\cite{Starobinsky:2001xq}) such as axion may lead to the flat spectra of 
isocurvature perturbations.  If the correlation between adiabatic and 
isocurvature perturbations is strong, adiabatic perturbations may be 
scale-invariant.

We wish to stress that our work is not conclusive. In particular, in 
order to fully evaluate the final power spectra, one should
solve the equations of motion for fluctuations in a 
nonsingular bouncing model including radiation.
Important issues which should be 
investigated further include:
\begin{itemize}
\item The final power spectra of the curvature perturbation
are found to be blue-tilted for the nonsingular Ekpyrotic models 
we have considered, which rely on specific higher derivative
correction terms. Are there other correction terms
to the action which are motivated by string theory, lead to 
nonsingular bouncing scenarios and yield a flat spectrum even in the 
single field case?  
Perhaps toy bouncing models using exotic scalar fields or matter
in refs.~\cite{Hwang:2001zt,Peter:2002cn} can be a good starting 
point to construct viable nonsingular Ekpyrotic models. 
\item We do not include the effect of radiation (or particles) which can be 
efficiently produced near the bounce.  Particle production around
the transition region is expected to be quite 
efficient \cite{Tsujikawa:2001ud}, and this
could lead to an additional instability of small-scale metric 
perturbations.  This effect may also 
non-trivially alter the nonsingular background evolution by the back-reaction 
effect of created particles.  It is also required to include the radiation 
after the bounce in order to evaluate the surviving spectra accurately, 
although the coupling between the scalar field and radiation should be 
chosen carefully in that case.
\item It is of interest to study 
the effect of isocurvature perturbations in the two-field system of 
nonsingular Ekpyrotic scenarios.  In particular isocurvature perturbations 
can be affected by the instability of the background near the bounce.  In 
order to obtain the final spectra of adiabatic perturbations, we need to solve 
the coupled equations of adiabatic and entropy perturbations through the 
nonsingular bounce including radiation.  It is important to investigate 
whether nearly scale-invariant spectra are obtained by conversion from 
isocurvature to adiabatic perturbations.
\end{itemize}

Our analysis  also applies to cyclic Universe models proposed in 
ref.~\cite{Steinhardt:2001vw,Steinhardt:2001st} in which the
bounce has been regularized by including higher-order corrections. Thus,
our conclusions about the difficulty in obtaining a scale-invariant
spectrum of fluctuations carry over to single field realizations of the cyclic
scenario.
In fact, we have done some simulations in the case of a simple 
negative potential $V=m^2(\phi^2-\phi_c^2)$ for $|\phi|<\phi_c$,
and found that the solutions can be nonsingular so long as
the higher-order effect dominates around the graceful exit.
Note, however, that the spectra of density perturbations
will be the same as in the Ekpyrotic scenario. 

Recently, a preprint has appeared \cite{Mukherji:2002ft} in which
in the context of a brane world scenario a nonsingular bouncing
cosmology is obtained by considering the motion of a D3 brane as a
boundary of a five dimensional charged anti de Sitter black hole.
In this model, computed in linear theory, the spectrum of 
gravitational wave fluctuations was shown not to be scale-invariant.
This result supports the conclusions we have reached \footnote{We
are grateful to the Referee for pointing out this reference.}.

\section*{APPENDIX~A: HEURISTIC DERIVATION OF THE SPECTRUM OF FLUCTUATIONS}

In this Appendix we give a heuristic derivation of the spectral index
of cosmological perturbations in the PBB and Ekpyrotic scenarios. This
analysis is based on two key assumptions. The first is
the assumption that the amplitude of the
fluctuations when they exit the Hubble radius during the phase of contraction
(in the Einstein frame) is given by the Hubble constant. This assumption
is reasonable assuming that the fluctuations are quantum vacuum perturbations
which freeze when their wavelength crosses the Hubble radius. 

The second assumption is that the ``physical magnitude'' of the fluctuations
remains unchanged while the wavelength of the fluctuation is larger than
the Hubble radius. This assumption is much less obvious, although at
first sight this assumption may seem obvious based on causality, namely the
fact that microphysics cannot influence physics on scales larger than
the Hubble radius. However, in inflationary cosmology and in models
with a contracting period such as the PBB and Ekpyrotic scenarios, the
forward light cone (causal horizon) is much larger than the Hubble radius, 
and the spatial coherence of background fields over scales of the forward
light cone can lead to nontrivial effects on fluctuation modes on these
scales, one of the most dramatic manifestations of this effect being
the parametric amplification of super-Hubble (but sub-horizon) cosmological
fluctuations during reheating in certain two-field inflationary models
\cite{Taruya:1997iv}-\cite{Tsujikawa:2002nf}.
Furthermore, the term ``physical magnitude'' of cosmological fluctuations 
is not well determined.  On super-Hubble scales, the magnitude of the 
density fluctuations depends sensitively on the coordinate system chosen.  
It is possible to choose coordinate-invariant (gauge-invariant) variables 
to describe the fluctuations, but there are many choices, and even in 
single field inflationary models many of these gauge-invariant fluctuation 
variables increase on super-Hubble scales (however, the increase between 
initial Hubble radius crossing during inflation at $t_i(k)$ and final 
Hubble radius crossing during the late time FRW cosmology at $t_f(k)$ is by 
a factor which only depends on the ratio of the equations of state at the 
two Hubble radius crossings).  This increase is a self-gravitational 
effect.

In spite of the above caveats, let us proceed with the heuristic
discussion of the amplitude of density fluctuations, applying it
first to inflationary cosmology (exponential expansion to be
specific). The quantity we wish to calculate is the mean square
mass fluctuation on a scale $k$ when the corresponding wavelength
enters the Hubble radius at final Hubble radius crossing $t_f(k)$.
This quantity, denoted $|{{\delta M} \over M}(k, t_f(k))|^2$ is
given by the power spectrum of fluctuations [see eq.~(\ref{PR})], and its
$k$-dependence on the spectral index $n$ is given by
\begin{equation}
\left|{{\delta M} \over M}(k, t_f(k)) \right|^2 \sim k^{n-1} \,.
\end{equation}  
Based on the first assumption,
\begin{equation}
\left|{{\delta M} \over M}(k, t_i(k))  \right|^2 \sim H^2(t_i(k)) 
\sim {\rm const.} \,,
\end{equation}
and using the second assumption we infer that
\begin{equation}
\left|{{\delta M} \over M}(k, t_f(k)) \right|^2 
\sim \left|{{\delta M} \over M}(k, t_i(k)) \right|^2 
\sim {\rm const.} \, ,
\end{equation}
and that hence the power spectrum is scale-invariant with an index $n = 1$.

PBB cosmology is characterized (in the Einstein frame) by a scale factor
which scales as
\begin{equation}
a(t) \sim t^{1/3}\,,
\end{equation}
and thus
\begin{equation}
H(t) = {1 \over {3 t}} \,.
\end{equation}
The condition of the initial Hubble radius crossing during the
period of contraction 
\begin{equation}
k a^{-1}(t_i(k)) = H(t_i(k))\,,
\end{equation}
leads to
\begin{equation}
t_i(k) \sim k^{-3/2}\,,~~~
H(t_i(k)) \sim k^{3/2}\,,
\end{equation}
and thus, applying our two basic assumptions as in the case of inflationary
cosmology, to
\begin{equation}
\left|{{\delta M} \over M}(k, t_f(k)) \right|^2 \sim k^3\,,
\end{equation}
which corresponds to a blue spectrum with index $n = 4$.

The analysis for Ekpyrotic cosmology is analogous. The only difference
is that the value of $p$ is different, $0 < p \ll 1$ and hence
\begin{equation}
t_i(k) \sim k^{- 1/(1-p)}\,,~~~
H(t_i(k)) \sim k^{1/(1-p)} \,,
\end{equation} 
and thus, taking $p = 0$ at the end,
\begin{equation}
\left| {{\delta M} \over M}(k, t_f(k)) \right|^2 \sim k^2 \,,
\end{equation}
which corresponds to a blue spectrum with index $n = 3$.

Obviously, given the caveats discussed at the beginning of this Appendix,
the results for PBB and Ekpyrotic cosmology cannot be trusted without
a fully relativistic analysis. The growth rates of cosmological
fluctuations are very different in expanding and contracting cosmologies,
and thus even given that the above heuristic analysis works in the
case of inflationary cosmology, this does not mean it has to work for
PBB and Ekpyrotic cosmologies. However, the results of our paper
are in agreement with those derived from the heuristic analysis.

\section*{APPENDIX~B: Analytic estimates for the gravitational potential}

Let us analyze the gravitational $\Phi$ in more details.  In the Einstein 
frame, the gravitational potential $\Phi_k^E$ in the longitudinal gauge is 
expressed in terms of ${\cal R}_k^E$ in the absence of 
higher-order corrections \cite{Finelli:1998bu,Hwang:2002fp}: 
\begin{eqnarray} 
\Phi_k^E=\frac{a_E^2\dot{H}_E}{k^2 H_E} \dot{R}_k^{E}\,.
\label{Phi}
\end{eqnarray}
The gravitational potential $\Phi_k^S$ in the string frame is
related with the one in the Einstein frame as \cite{Hwang:re} 
\begin{eqnarray} 
\Phi_k^S=\Phi_k^E+\frac{\delta F}{2F}=\Phi_k^E-\frac12 \delta\phi_k \,.
\label{rel}
\end{eqnarray}
Making use of eqs.~(\ref{ta}) and (\ref{relation}), we  find that 
the curvature perturbation in the Einstein frame is exactly the same as 
that in the string frame (i.e, ${\cal R}_k^E={\cal R}_k^S$).  
Therefore the gravitational potential in the Einstein frame is expressed in 
terms of $\dot{{\cal R}}_k^S$: 
\begin{eqnarray} 
\Phi_k^E=\frac{a_S^2\left( \dot{H}_S-\ddot{\phi}/2
+ \dot{\phi}H_S/2-\dot{\phi}^4/4\right)} {k^2(H_S-
\dot{\phi}/2)} \dot{\cal R}_k^S\,.
\label{modi}
\end{eqnarray}
Note that dots in eq.~(\ref{modi}) denote the time-derivative with respect 
to $t_S$.  This is the equation which we solve numerically.

Taking note of the relation, $H_{\nu}'(x)=H_{\nu-1}(x)-(\nu/x)H_{\nu}(x)$, 
one finds 
\begin{eqnarray} 
 \dot{\cal R}_k^S=\frac{\sqrt{\pi}}{4a_Sz} \left[ 2\sqrt{s|\eta_S|}k 
 \left(c_1 H_{\nu-1}^{(1)}(x) +c_2 H_{\nu-1}^{(2)}(x) \right) \pm 
 \frac{1}{\sqrt{|\eta_S|}}\left(1-2\nu \mp 2|\eta_S| \frac{z'}{z}\right) 
 \left(c_1 H_{\nu}^{(1)}(x) +c_2 H_{\nu}^{(2)}(x) \right) \right]\,,
\label{dotPsi}
\end{eqnarray}
where each sign corresponds to the case with $\eta_S>0$ and $\eta_S<0$,
respectively. 
When the evolution of $z$ is given by $z \propto |\eta_S|^{\gamma}$,
we have
\begin{eqnarray} 
1-2\nu \mp 2|\eta_S| \frac{z'}{z} =1-2\nu- 2\gamma
=1-2\gamma-\left|1-2\gamma \right|=
\cases{ 0 & for $\gamma<1/2$ \,, 
\cr 2(1-2\gamma) & for $\gamma>1/2$ \,.  \cr }
\label{re}
\end{eqnarray}
This term completely vanishes during the contracting  phase in the 
Ekyprotic cosmology with $p<1/3$, since $\gamma$ is less than $1/2$.
In this case the gravitational potential in the Einstein frame can be 
expressed as
\begin{eqnarray} 
\Phi_k^E=\frac{a_S \sqrt{\pi s |\eta_S|}\left( \dot{H}_S-\ddot{\phi}/2
+ \dot{\phi}H_S/2-\dot{\phi}^4/4\right)}{2z(H_S-\dot{\phi}/2)k} 
\left(c_1 H_{\nu-1}^{(1)}(x) +c_2 H_{\nu-1}^{(2)}(x) \right)\,.
\label{Phi3}
\end{eqnarray}
This relation is used to estimate the spectrum of $\Phi_k^E$
in Sec.~IV.

\section*{ACKNOWLEDGMENTS}
We are grateful to Jai-chan Hwang, Daisuke Ida, Justin Khoury, Jerome Martin, 
Patrick Peter, Alexey Toporensky, Gabriele Veneziano, David Wands
and Jun'ichi Yokoyama for useful discussions.  
F.F.  wishes to thank Cyril Cartier, 
Ruth Durrer and Filippo Vernizzi for their invitation to visit the University 
of Geneva and for stimulating discussions.  S.T.  is grateful for financial 
support from the JSPS (No.  04942).  R.B.  wishes to thank the CERN Theory 
Division and the Institut d'Astrophysique de Paris for their hospitality 
and support during the time the work on this project was done.  He also 
acknowledges partial support from the US Department of Energy under 
Contract DE-FG02-91ER40688, TASK A.  
Work similar to the study reported on here is being
carried out by C. Cartier, R. Durrer and F. Vernizzi.


\end{document}